\documentclass[aps,prb,reprint,groupedaddress]{revtex4-2}
\usepackage{graphicx}
\usepackage{bm}
\usepackage{xcolor}

\begin{document}

\newcommand{\be}{\begin{equation}}
\newcommand{\ee}[1]{\label{#1}\end{equation}}
\newcommand{\bem}{\begin{eqnarray}}
\newcommand{\eem}[1]{\label{#1}\end{eqnarray}}
\newcommand{\eq}[1]{Eq.~(\ref{#1})}
\newcommand{\Eq}[1]{Equation~(\ref{#1})}
\newcommand{\ua}{\uparrow}
\newcommand{\da}{\downarrow}
\newcommand{\g}{\dagger}
\newcommand{\rc}[1]{\textcolor{red}{#1}}

% Use the \preprint command to place your local institutional report
% number in the upper righthand corner of the title page in preprint mode.
% Multiple \preprint commands are allowed.
% Use the 'preprintnumbers' class option to override journal defaults
% to display numbers if necessary
%\preprint{}

%Title of paper
\title{Quantum rotator and  Josephson junction:  compact vs. extended  phase \\
and dissipative quantum phase transition}

% repeat the \author .. \affiliation  etc. as needed
% \email, \thanks, \homepage, \altaffiliation all apply to the current
% author. Explanatory text should go in the []'s, actual e-mail
% address or url should go in the {}'s for \email and \homepage.
% Please use the appropriate macro foreach each type of information

% \affiliation command applies to all authors since the last
% \affiliation command. The \affiliation command should follow the
% other information
% \affiliation can be followed by \email, \homepage, \thanks as well.
%\homepage[]{Your web page}
%\thanks{}
%\altaffiliation{}

\author{Edouard  B. Sonin}
\email[]{sonin@cc.huji.ac.il}

%\homepage[]{Your web page}
%\thanks{}
%\altaffiliation{}
\affiliation{Racah Institute of Physics, Hebrew University of Jerusalem, Givat Ram, Jerusalem 9190401, Israel}

%Collaboration name if desired (requires use of superscriptaddress
%option in \documentclass). \noaffiliation is required (may also be
%used with the \author command).
%\collaboration can be followed by \email, \homepage, \thanks as well.
%\collaboration{}
%\noaffiliation

\date{\today}

\begin{abstract}
The paper reassesses the old dilemma  ``compact vs. extended  phase'' in the  quantum theory of the rotator and the Josephson junction and the analogy of these two systems with the  a particle moving in a periodic potential. This dilemma is in fact the dilemma whether the states with  the phases $\varphi$ and $\varphi+2\pi$ are distinguishable, or not. In the past it was widely accepted that in the Josephson junction these states are distinguishable like for  a particle moving in a periodic potential. This paper argues that  the states with  the phases $\varphi$ and $\varphi+2\pi$ are indistinguishable as in a pendulum (a particular example of the quantum rotator). However, this does not lead to revision of the {\em conclusions} of the conventional theory predicting the transition from the superconducting to the insulating state in the small    Josephson junction.

\end{abstract}

% insert suggested keywords - APS authors don't need to do this
%\keywords{}

%\maketitle must follow title, authors, abstract, and keywords
\maketitle

\section{Introduction}

The paper is written for the special issue dedicated to the outstanding physicist Mark Azbel. It addresses the problems of quantum physics, to which Prof. Azbel made a number of seminal 
theoretical contributions \cite{AzbelP,AzbelN,Azbel}. 

The angle variable $\varphi$ is ubiquitous in classical  and  quantum physics. Among  examples of this variable are the rotation angle of the plane rotator  (mechanical pendulum  as a particular case) and the phase difference across  the single Josephson junction (JJ).  The  canonically  conjugate variable to the phase is the angular momentum $M$ (further called moment) in the first example and  the charge  in the second one. 

The history of using the canonically conjugate  pair  ``angle  (phase)--moment'' in quantum mechanics is full of controversies and disputes. In particular, the commutation  relation 
\be
[\hat \varphi,\hat M]=i\hbar
    \ee{ComS}
introduced by \citet{dirac} was challenged \cite{judge,PhasRev,Pegg}. Here  $\hat \varphi$ and $\hat  M$ are operators of the angle  (phase) and  the moment, respectively. The problem with the commutation relation was connected with  the non-Hermitian character  of  the phase operator. It was suggested to repair this mostly mathematical flaw  by rewriting the commutation relation in the terms of Hermitian operators  $\sin \hat \varphi$ and $\cos \hat \varphi$ \cite{suss}. The uncertainty relation 
\be
\Delta M \Delta \varphi > {\hbar\over 2} 
   \ee{unc}
was also under scrutiny \cite{judge,PhasRev,Pegg}.  Here  $\Delta M$ and $ \Delta \varphi$ are uncertainties of the moment and the phase, respectively.
 
Another problem  (but connected with the first one) is that the phase  $\varphi$ is defined modulo $2\pi$. One can choose the phase defined in  an   interval from an arbitrarily chosen phase  $\varphi_0$  to $\varphi_0  +2\pi$  (compact phase),  or  the phase ranging from $-\infty$ to $\infty$ (extended phase). If  the  phases differing by the integer number $s$ of  $2\pi$ describe the same state, it does not matter  at all which phase, compact or extended,  one uses in  the theoretical analysis.  The analysis (if correct) must lead to the same results. However, in quantum mechanics there are some  subtleties, and there is no consensus  on the dilemma ``compact vs. extended phase''. 

The proponents of  the suggestion that only the compact phase must be used in the quantum theory of the JJ argue that it is natural to expect that that the states with the phases $\varphi$ and $\varphi+2\pi$ are the same state and  only states with wave functions periodic in  $\varphi$ with the period $2\pi$ are possible. This means that the variable canonically conjugated to the phase (moment in a quantum rotator, or charge in a JJ) is quantized. The proponents of the extended  phase argue, that the ``natural expectation'' of the identity of the states with the phases $\varphi$ and $\varphi+2\pi$ is not so natural and is invalid in the case of the JJ because of its interaction with the environment.  Then different phases in the whole interval  $-\infty<\varphi <\infty$ always correspond to different states. This was called ``decompactification of the phase''.  

It  is important to stress, however, that the compact phase is sufficient for description of states, but not for description of dynamical processes of transitions between states with different phases. In these processes it is necessary  to know  not only the variation of the compact phase but also an integer number $s$ of full $2\pi$ winding during the process.  It is more convenient instead of two variables to deal only with one variable, which is an extended phase determined in the interval $(-\infty,\infty)$
\be
\varphi(t)  =2\pi s(t) +  \varphi_c(t).
      \ee{comp}
Here $\varphi_c$ is the compact phase determined in any interval of the length $2\pi$. The voltage drop over the JJ is determined by the time derivative of the extended but not the compact phase. The time derivative of the compact phase has unphysical jumps when the phase reaches borders of the $2\pi$ interval chosen for the compact phase.
Thus, one should not interpret  the requirement of using only the compact phase (phase compactification) literally but interpret it as  the requirement of using only wave functions periodic in the extended phase $\varphi$ with the period $2\pi$. Under  the assumption of decompactification this requirement is abandoned.
Thus, the dilemma ``compact vs. extended  phase'' is in fact  the dilemma whether the  states with the phases $\varphi$ and $\varphi+2\pi$  indistinguishable or distinguishable. Nevertheless, further in the paper the dilemma will be called  ``compact vs. extended  phase'' as widely accepted in the literature.

Actuality of this dilemma  reemerged in the recent dispute about the Dissipative Quantum Phase  Transition (DQPT) \cite{Sacl,CommMurani,ReplMurani}  between the superconducting and insulating states of a single JJ predicted about 40 years ago \cite{Schmid,Bulg}.  \citet{Sacl} claimed that their  experiment and theory proved the absence of  the  DQPT, because the single JJ cannot be an insulator. 

The estimation done in Ref.  \onlinecite{CommMurani} demonstrated that the experiment of Murani {\em et al.} was done at the conditions, in which the existing theory did not predict the DQPT. Therefore, the experiment could not provide  any evidence either pro or contra the DQPT. Their theoretical arguments were also rejected, but they deserve an  analysis more detailed than it was possible within  a short Comment \cite{CommMurani}. In particular, Murani {\em et al.} \cite{Sacl,ReplMurani} argued that the conventional theory failed because it  used the extended phase while  only the compact phase must be used. This bring us to the topic of the present paper.

Because of generality of the aforementioned problems with the phase variable, the paper addresses three systems: quantum rotator, particle in a periodic potential, and single JJ.  In the quantum rotator a particle moves around a 1D ring. The quantum pendulum \cite{Cond} and an electron moving around a 1D normal ring \cite{silver,AzbelN,Buttiker,Gefen} are examples of the quantum rotator. The paper explores analogies between these systems, but looking for possible differences at the same time. 

In order to resolve the dilemma ``compact vs. extended  phase'',  it is necessary to answer to three questions:
\begin{enumerate}
\item
Are the states with the phases $\varphi$ and $\varphi+2\pi$ indistinguishable in the JJ?
\item
Must the wave function be periodic in $\varphi$ if the states with the phases $\varphi$ and $\varphi+2\pi$ are indistinguishable?
\item
The last but not the least: Is it important for the theory of DQPT whether the states with the phases $\varphi$ and $\varphi+2\pi$ are  distinguishable, or not?
\end{enumerate}
The paper looks for answers to these three questions.

Our analysis has fully confirmed the final conclusions of the about 40-years old conventional theory of the  DQPT in the small JJ. But it reassessed  justifications of  these conclusions and analogies of the single JJ with other systems  (quantum rotator and particle in a periodic potential).  While in the past it was widely (but not unanimously) believed that the single JJ is an analog of particle in a periodic potential, but not of a mechanical pendulum, we argue that the opposite is true. This means that the states with the phases $\varphi$ and $\varphi+2\pi$ are distinguishable both in the JJ and the quantum rotator.  However, whatever analogy is more correct, the  DQPT must take place both in the JJ and  in the quantum rotator.

\section{The phase in the classical theory} \label{cl}

\subsection{Plane rotator in the conjugate variable ``angle--moment''}            \label{QRc}

The Hamiltonian of the classical plane rotator  is 
\be
H={m^2\over 2 J}+G (1-\cos \varphi )-\varphi N,
      \ee{Ho}
where $m$ is the moment of the particle moving in the rotator, $J$ is  the moment of  inertia, $N$ is the external torque, and the periodic in $\varphi$ potential $\propto G$
emerges from a constant force acting on the rotator (the gravity force in the case of the pendulum or the constant  electric field in the case of a charged particle  in a 1D ring). The $\varphi$-dependent part of the Hamiltonian is the well known washboard potential.
The Hamilton equations are
\bem
{dm\over dt}= -{\delta H\over \delta \varphi }=- G\sin\varphi +N,
 \nonumber \\
 {d\varphi \over dt}={\delta H\over \delta m}={m\over J}.
   \eem{phit}

The Hamiltonian \eq{Ho} is not periodic in $\varphi$. This looks as a flaw, since violates the principle that the states with the phases $\varphi$  and $\varphi+2\pi $ are indistinguishable. 
According to this Hamiltonian, the energies of these states differ by the energy $2\pi N$ pumped to the rotator from the environment after full $2\pi$ winding of the rotator.

The flaw can be eliminated by using another Hamiltonian 
\be
H={( M+M_0)^2\over 2 J}+G (1-\cos \varphi ), 
            \ee{Hm}
which is periodic  in $\varphi$. Here  the moment $M_0$ transferred to the rotator by  the external  torque,
\be
{dM_0\over dt}=N,
    \ee{}
was introduced. Since $M_0$ emerges from the interaction with the environment, we shall call it {\em external moment}. 
The Hamilton equations for the Hamiltonian \eq{Hm} are

\bem
{d M\over dt}= -{\delta H\over \delta \varphi }=- G\sin\varphi ,
\nonumber \\
{d\varphi \over dt}={\delta H\over   \delta  M}={ M+M_0\over J}.
   \eem{phgP}
The moment  
\be
 M ={\partial {\cal L}\over \partial \dot \varphi}
    \ee{}
is the canonical moment determined by the Lagrangian 
\be
{\cal L}={(J \dot \varphi -M_0)^2\over 2 J}-G (1-\cos \varphi ).
    \ee{}
However, the angular  velocity $\omega=\dot \varphi$ of the phase rotation is determined not by the canonical  but by the kinetic moment $m=  M+M_0$. The equations of motion \eq{phit}  in  the terms of the observables $m$ and $\varphi$ do not depend on the choice of  the Hamiltonians \eq{Ho}     or \eq{Hm}, since they differ by the full time derivative from $\varphi M_0$, which does not affect the equations of motion. Later on we shall call the gauges with the periodic Hamiltonian like in \eq{Hm} and with the non-periodic Hamiltonian like in \eq{Ho} {\em periodic }and {\em non-periodic gauge}, respectively.

The terms  {\em canonical moment} and {\em kinetic moment} were introduced by the analogy with the {\em canonical momentum} and {\em kinetic momentum}  of a charged particle in the electromagnetic field.  Splitting of the kinetic moment $m$ onto the canonical and the external moment obtained  from the environment is purely formal in the classical theory. But this splitting is more important in the quantum theory. 

 In the  absence of the torque   $ N$, which pumps the moment and the energy to the system, there are two types of motion: (i) an oscillation around the ground state $\varphi=0$ with $m$ vanishing  in average, and (ii) a monotonic rotation with $\langle m \rangle \neq 0$  $\varphi(t)$ being a periodic function in the time interval from $-\infty$ to $\infty$. 
The stationary state with constant 
\be
\varphi =\arcsin(N/G)+2\pi s
     \ee{}
  is possible if $N <G$.   In this state the kinetic moment $m$ vanishes.  The  states of the rotator at the phases $\varphi$ and $\varphi+2\pi s$ are indistinguishable.  The stationary state with   constant $\varphi $  at  $N <G$ is an analog of the superconducting state of the JJ, while the regime of monotonic rotation  is an analog of the insulating state of the JJ.  At  $N >G$  the torque  drives the  quantum rotator to rotate with acceleration, but the monotonic  rotation with the angular velocity $\omega=d\varphi /dt$ periodically oscillating around the constant average angular velocity $\bar \omega$  is possible  in the presence of  friction.

\subsection{Plane rotator vs. particle in a periodic potential}

Let us  compare the rotator dynamics  with the dynamics of a particle with charge $q$ moving in a periodic potential and a classical electromagnetic field. In the latter case the Hamiltonian is 

\be
H={( P- {q\over c} A)^2\over 2 m_0 }+G\left(1-\cos {2\pi x\over l} \right) +q\Phi. 
            \ee{Hg}
Here $m_0$ is the particle mass, $c$ is the speed of light, $P$ and $A$ are  the $x$-components of the canonical momentum $\bm P$ and  of the electromagnetic vector potential $\bm A$ (we consider the 1D problem), and $\Phi$ is the electromagnetic scalar potential. The gauge transformation,
\be
A=A' +\nabla \chi(x,t),~~\Phi=\Phi' -\dot   \chi(x,t),
     \ee{GT}
with $\chi(x,t)$ being an arbitrary function of $x$ and $t$,
yields the Hamiltonian \eq{Hg} with $A'$ and $\Phi'$ instead of $A$ and $\Phi$ and with the full time derivative of $q\dot   \chi(x,t)$ added, which  does not affect the Hamilton equations of motion.

The torque on the plane rotator can be a result  of the magnetic field normal to the plane of the rotator. The Hamiltonian   \eq{Hg} describes also the dynamics of the plane rotator with $x$ being the coordinate along the circumference of the 1D ring of the rotator and $l$ being the length of this circumference.  The relations between variables in two presentations are 
\be
\varphi={2\pi x\over l} ,~~M={Pl\over 2\pi},~~J={m_0l^2\over 4\pi^2} .
    \ee{cord}
The periodic gauge in Sec.~\ref{QRc} corresponds to the gauge without the scalar potential $\Phi$ and with the Hamiltonian 
\be
H={( P- {q\over c} A)^2\over 2 m_0 }+G\left(1-\cos {2\pi x\over l} \right), 
            \ee{HmS}
The external moment and the external torque of Sec.~\ref{QRc} are
\be
M_0=-{ql \over 2\pi c }A=-\hbar{\phi\over \phi_0},~~N=\dot M_0={ ql  \over 2\pi}E,
    \ee{cor}
where $E=-\dot A/c$ is the azimuthal component of the electric field and $\phi_0=hc/q$ is the magnetic flux quantum  for the particle of charge  $q$.
The external  moment $M_0$  is determined by the  magnetic flux $\phi=Al$ through the area restricted by the 1D ring of the rotator. The magnetic field is supposed to be  axisymmetric.

The transformation with the gradient  $\nabla \chi = A(t)$ independent from $x$ yields the non-periodic gauge, in which the potential $A(t)$ is absent, but instead the linear in $x$ scalar potential $\Phi(t)=-  E(t) x=\dot A(t) x/c$ appears: 
\be
H={ P^2\over 2 m_0 }+G \left(1-\cos {2\pi x\over l} \right)- q E x. 
            \ee{WS}

The scalar potential in the non-periodic gauge is multivalued. 
This does not produce any problem, since only fields but not potentials are observable quantities.

 Whatever gauge one uses, there is no difference between  dynamics of the rotator and the charged particle in a periodic potential. The dynamics does not depend on whether  the positions of the particle with the coordinates $x$ and  $x+l$ (or angles $\varphi$ and $\varphi+ 2\pi $) are distinguishable,  or not.  Thus, in the classical theory the dilemma ``compact vs. extended  phase'' does not exist and there is no difference between the dynamics in a 1D ring and in the infinite 1D space with a periodic potential.

\section{The phase in the quantum  theory}

\subsection{Axisymmetric quantum  rotator: commutation and uncertainty relations, and wave packets} \label{WP}

The standard way to go from the classical to the quantum theory is to replace  in the Hamiltonian the canonical moment $ M$  by the operator
\be
\hat { M} =-i\hbar {\partial  \over \partial \varphi}.
    \ee{Mphi}

General problems  with the canonically conjugate  pair  ``angle  (phase)--moment'' can be discussed for
the simple case of  the axisymmetric rotator ($G=0$). This case has already been investigated in the works on persistent currents in 1D normal rings \cite{Buttiker,Gefen}. For a while we also ignore 
 the external moment $M_0$. This means that we ignore any interaction with the environment, either at the present moment, or in the past. 
 
 The objection to commutation relation \eq{ComS} was based on the following calculation of the matrix elements of the commutation relation between two eigenstates of the moment operators with eigenvalues $M_s$ and $M_{s'}$ \cite{judge,PhasRev,Pegg}:
\bem
\int\limits_0^{2\pi}\psi^*_{s'} [\hat \varphi,\hat M]\psi_s\,d\varphi
\nonumber \\
=-i\hbar \int\limits_0^{2\pi}\psi^*_{s'}\left[ \varphi{\partial \psi_s\over \partial \varphi}-{\partial (\varphi \psi_s)\over \partial \varphi} \right]\,d\varphi
\nonumber \\
=-i\hbar\int\limits_0^{2\pi}\left(\psi^*_{s'} \varphi{\partial \psi_s \over \partial \varphi}+{\partial \psi^*_{s'} \over \partial \varphi}\varphi \psi_s\right)\,d\varphi
 \nonumber \\
=\left. i\hbar \psi^*_{s'} \varphi\psi_s  \right|_0^{2\pi}+(M_s-M_{s'})\int\limits_0^{2\pi}\psi^*_{s'} \varphi\psi_s \,d\varphi
    \eem{ComC}
The opponents of the commutation relation \eq{ComS} neglected the first term emerging from the borders of the integration interval $(0,2\pi)$.  Then the diagonal matrix elements ($M_s=M_{s'}$) of the commutator  vanish, while the diagonal matrix elements of the righthand side of the commutation relation \eq{ComS} are not zero. The justification
for ignoring of the border contribution  was that it should not appear in a matrix element of a Hermitian  operator when the integrand is a periodic function of $\varphi$. 
The operator  $\hat \varphi$ in the commutation relation is not Hermitian and breaks periodicity. 

Various modifications of the commutation relation were suggested (one of them is discussed below).  However, there is another resolution of the problem, which rehabilitates the commutation relation \eq{ComS}. The border  term in \eq{ComC}  appears after the integration by parts of only one from two terms in the original commutator. While any of two terms in the commutator separately is non-Hermitian and breaks periodicity, their sum is  Hermitian and does not break periodicity. The matrix element of the commutator can be calculated without integration by parts:
\bem
\int\limits_0^{2\pi}\psi^*_{s'} [\hat \varphi,\hat M]\psi_s\,d\varphi
\nonumber \\
=-i\hbar \int\limits_0^{2\pi}\psi^*_{s'}\left( \varphi{\partial \psi_s\over \partial \varphi}-\varphi {\partial \psi_s\over \partial \varphi} -\psi_s\right)\,d\varphi
\nonumber \\
=i\hbar \int\limits_0^{2\pi}\psi^*_{s'}\psi_s\,d\varphi.
    \eem{ComP}
This is equal to the matrix element of the righthand side of the commutation relation. Similar arguments rehabilitating the standard commutation relation \eq{ComS} were presented by \citet{LossM}.

We checked the commutation relation using the wave functions in the continuous space of the phase $\varphi$. Another route is to do it in the discrete space of quantized moments $M$. Then one encounters the  problem that because of discreteness of $M$ the expression conjugate to \eq{Mphi}
\be
\hat \varphi =i\hbar {\partial  \over \partial M}
    \ee{phiM}
for the operator $\hat \varphi$ in the moment space is invalid. Instead one can use the operator $e^{i\hat \varphi}$, which shifts from one quantized eigenvalue of the operator $\hat M$   to the next  one.
The commutation relation with this operator is
\be
[e^{i\hat \varphi},\hat M]=-\hbar e^{i\hat \varphi}.
    \ee{eph}

The operator $e^{i\hat \varphi}$ is a superposition of two Hermitian operators $\cos \hat \varphi$ and $\sin \hat \varphi$. The commutation relations for these operators, which are equivalent  to \eq{eph}, were suggested by \citet{suss}. Although the commutation relation \eq{eph} contains only Hermitian  operators, its expansion in $\hat \varphi$ consists of non-Hermitian  operators, which must be treated correspondingly. A failure of some operation valid only for Hermitian  operators, means the failure of the operation, but not of the commutation relation.

The problem with the canonical commutation relation naturally leads to the problem with the uncertainty relation \eq{unc}. The uncertainty relation is derived from the analysis of semiclassical wave packets, which demonstrates the correspondence principle: the transition from the quantum mechanical to the classical description. The wave packet is formed by a superposition of the states with  moments $M$ in the interval of the width $\Delta M$. In the continuous moment space the superposition is determined by an integral. In the discrete moment space the integral must be replaced by a sum over the quantized values of the moment.

 Let us assume that the phase uncertainty is essentially less than $2\pi$.  Then the moment uncertainty
\be
\Delta M \gg {\hbar\over 4\pi} 
   \ee{}
is much larger than the distance between quantized values of the moment. Then the quantization can be ignored, and the summation determining the wave packet can be replaced by the integration. This provides a sufficiently accurate description of the wave packet within the phase interval $2\pi$. 

But there is a hurdle in this picture. The original wave function of the wave packet with summation over quantized values of the moment is periodic in $\varphi $ because any term of the sum is periodic. However, the replacing of summation by integration definitely breaks the periodicity. The flaw is easily healed. The wave packet is replaced by the periodic chain of packets. The procedure can be considered as a compactification of phase, which the proponents of the compact phase insist on. Namely, one calculates the phase only in one $2\pi$ interval and then continues this wave function periodically on all other phase intervals.  At the same time this demonstrates that the compactification is necessary only due to inaccuracy  of the approximation and is not needed for a sufficiently exact analysis.

Another situation emerges if one tries to construct a wave packet with the phase uncertainty $ \Delta \varphi$ exceeding $2\pi$.  In this case the moment uncertainty $\Delta M <\hbar/4\pi$ is less than the moment quantum, the summation reduces to only one term, and the picture of wave packets fails. Then the physical meaning of the uncertainty relation \eq{unc} becomes unclear.

There were suggestions to modify the standard uncertainty relation \eq{unc} as a reaction to the aforementioned problem \cite{judge,Pegg}. They were based on the concept of the compact phase, which assumed that the phase uncertainty cannot exceed $2\pi$. This concept ignores that a phase fluctuation cannot be described only by the fluctuation of the compact phase (Sec.~\ref{cl}) . A number $s$ of full $2\pi$ rotations [see \eq{comp}] in the course of the fluctuation is also important. We do not dwell more on this issue, since our analysis of the slow dynamics is based on the adiabatic approximation and does not use the wave packet concept.

\subsection{Particle moving in a 1D ring vs. particle moving in an infinite 1D space} \label{per}

Let us now compare the quantum mechanical dynamics of  a particle moving  in a 1D ring of the rotator and a  particle moving in the infinite 1D space. As in the previous subsection,
we ignore the periodic potential $G\left(1-\cos {2\pi x\over l} \right)$. This allows to deal with simple analytical solutions of the Schr\"odinger equation.

In the quantum mechanics the canonical momentum becomes an operator:
\be
\hat P=-i\hbar {\partial \over \partial x}.
     \ee{}
The quantum mechanical version of the Hamiltonian  \eq{Hg} and the Schr\"odinger equation at $G=0$ in the periodic gauge  are 
\be
H={1\over 2 m_0}\left|-i \hbar{\partial \psi \over \partial x} -{q\over c}A(t)\psi\right|^2, 
            \ee{Hcx}
\be
i\hbar {\partial \psi \over \partial t}=-{1\over 2m_0}\left[ \hbar{\partial  \over \partial x} -i{q\over c}A(t)\right]^2\psi  .
            \ee{Shcx}
The Schr\"odinger equation  has a solution for an arbitrary time dependence of  $A(t)$:
\be
\psi(\varphi,t) = e^{iP x /\hbar-i \int^t{{\cal E}(t')\over \hbar}dt'},
    \ee{wfc}
which is an eigenstate on the canonical momentum with the  eigenvalue $P$.  Here the time dependent energy is
\be
{\cal E}(t)={[P -{q\over c}A(t)]^2\over 2m_0}.
     \ee{}
The particle  velocity 
\be
v ={dx\over dt} ={d{\cal E}\over d P}={P -{q\over c}A(t)\over m_0}
    \ee{omega0}
depends on the kinetic momentum $p=P -{q\over c}A(t)$ and is well-defined, while the coordinate $x$ itself is not defined at all. There is an equal probability for any value of $x$.  
An electric field  $E=-\dot A/c$ monotonically accelerates the particle, as in the classical theory.

In a constant electric field
\be
\psi(\varphi,t) = e^{iP x /\hbar- {i(P+qEt)^3\over 6m_0 qE}}.
    \ee{}

\begin{figure}[!b]
\includegraphics[width=0.4 \textwidth]{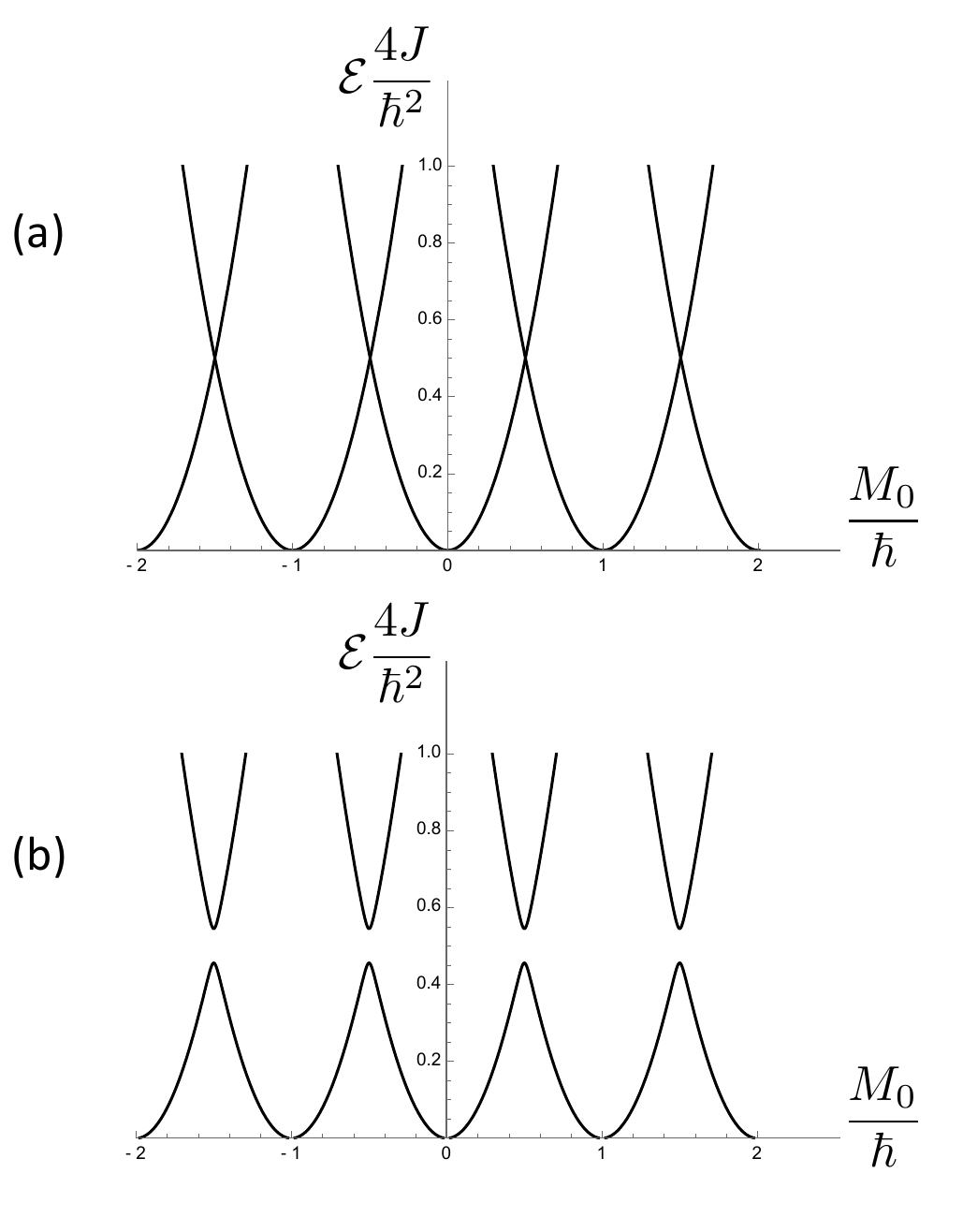} 
 \caption{ Plot the energy vs. the external moment $M_0$ at various quantized values of the canonical moment $M=s\hbar$. (a) Axisymmetric rotator.  (b) Quantum rotator in a constant field.      \label{f1}}
 \end{figure}

In the quantum mechanics  the difference of the particle dynamics in the 1D infinite space  and in the 1D ring  becomes  important. In the former case any value of $P$ is allowed. In the latter case  
the canonical momentum $P$ is quantized and cannot differ from the values $sh/l$ with integer $s$. Only at these quantized  values the wave function \eq{wfc} is periodic with the period $l$.  
In the variables ``moment--angle'' the quantized values of the canonical moment $M=Pl/2\pi$ [see \eq{cord}] are $s\hbar$.
The plot the energy vs. the external moment $M_0$ for different quantized values of the canonical moment  is shown in Fig.~\ref{f1}(a).

In the quantum theory the gauge transformation \eq{GT} must be accompanied by the transformation of the wave function \cite{LLqu}:
\be 
\psi =\psi'e^{iq\chi/\hbar c}.
      \ee{GTp}
After the transformation with $\chi =xA=-xc\int^tE(t')dt'$,
\be 
\psi =\psi'e^{i qAx/\hbar c},
      \ee{gauge}
  from the periodic to the non-periodic gauge, the Hamiltonian and the  Schr\"odinger equation become
\be
H={\hbar^2\over 2 m_0}\left|{\partial \psi' \over \partial x}\right|^2- qE(t)x|\psi'|^2,
            \ee{Hwb0}
\be
i\hbar {\partial   \psi' \over \partial t}=-{\hbar^2\over 2m_0}{\partial^2 \psi '\over \partial x^2} - qE(t)x\psi'.
   \ee{Sht0}
The gauge transformation  \eq{gauge} yields an non-stationary state with the non-periodic wave function
 \bem
\psi'=\psi e^{-qAx/c} = e^{i\left(P +qE t\right) x /\hbar-i \int^t{{\cal E}(t')\over \hbar}dt'}.
    \eem{nonpf}
 In a constant electric field
\be
\psi'(\varphi,t) = e^{i(P+qEt) x /\hbar- {i(P+qEt)^3\over 6m_0 qE}}.
    \ee{}

After the gauge transformation the canonical and the  kinetic momentum do not differ and are determined by the operator
\be
\hat p'=\hat P'=-{\partial \over \partial x}
    \ee{}   
in the space of functions $\psi'$. In the quantum rotator  the quantization of the canonical momentum must be done in the periodic but not in the non-periodic gauge. This means that not the momentum $p'$ but the canonical momentum $P=sh/l$  is equal  to an integer number $s$ of the momentum quanta. 

In the non-periodic gauge the wave function is non-periodic since the gauge transformation \eq{GTp} and the Hamiltonian \eq{Hwb0} are non-periodic. One can rewrite  \eq{nonpf} as 
\bem
\psi'= e^{iPx /\hbar-i \int^t{{\cal E}'(t')\over \hbar}dt'},
    \eem{}
where 
\be
{\cal E}'(t)={\left(P +qEt\right)^2\over 2J}-qEx
     \ee{}
is the energy after the  gauge transformation. It is evident that the  wave function is not periodic because of the non-periodic term $-qEx$ in the energy.  It is impossible to satisfy the requirement of the wave function periodicity in the non-periodic gauge. If the wave function  is periodic at some moment of time it will become non-periodic at the next moment of time because of the non-periodic Schr\"odinger equation.

The loss of periodicity of the wave function in the non-periodic gauge should not be a matter of concern, as well as not  a matter of concern is the non-periodic electric scalar potential.  The phase factor, which makes the wave functions $\psi(x)$ and $\psi(x+l)$ different, means that the state is described by a multivalued wave function. The property of the gauge transformation to make the wave  function multivalued was pointed out by \citet{LLqu} in Sec. 111 of their book. Multivaluedness (non-periodicity) of the wave function of the quantum rotator compensates multivaluedness (non-periodicity) of the washboard potential  in the non-periodic gauge \cite{david}.

In summary, the important and the only difference between the dynamics of the particle in the quantum rotator and the particle moving in the infinite 1D space is that the Hilbert space of wave functions in the former case is discrete and is a subspace of the continuous Hilbert state in the latter case.

\subsection{Quantum  rotator in an external constant field} \label{rcf}

While in Sec.~\ref{per} the variables ``coordinate--momentum'' were more convenient for comparison of the rotator with the particle moving in the infinite 1D space, here we return back to the variables ``angle (phase)--moment'', which are more convenient for comparison with the JJ.

At the presence of the external constant field the quantum mechanical version of  the  Hamiltonian \eq{Hm} in the periodic gauge is
\be
H={1\over 2 J}\left|-i \hbar{\partial \psi \over \partial \varphi} +M_0\psi\right|^2+G (1-\cos \varphi ) |\psi|^2.
            \ee{Hms}
The Schr\"odinger equation for this Hamiltonian is
\be
i\hbar {\partial \psi \over \partial t}=-{1\over 2J}\left(\hbar{\partial \over \partial \varphi}+iM_0\right)^2\psi +G (1-\cos \varphi ) \psi.
   \ee{Sh}

According to  the Bloch theorem,  any stationary  solution of \eq{Sh} is the Bloch function
\be
\psi(\varphi,t)=u(\tilde M+M_0,\varphi)e^{i (\tilde M \varphi- E_0 t)/\hbar},
   \ee{BF}
where $u(\tilde M+M_0,\varphi)=u(\tilde M+M_0,\varphi+2\pi)$ is a periodic in  $\varphi$  function and   $\tilde M$ is a canonical quasimoment (analog of the canonical quasimomentum in the solid body theory \cite{LLstPh2}). The energy spectrum of Bloch states consists of bands  with forbidden  gaps between them. In the quantum rotator the  Bloch wave function must be periodic in $\varphi$. It is possible only  for quantized values  of the canonical quasimoment  $\tilde M=s\hbar$.

We consider only the lowest band with the energy $E_0(\tilde M  +M_0)$, which depends on  the kinetic quasimoment  $\tilde m=\tilde M  +M_0$. 
By the analogy with the quasimomentum and the coordinate operators for a particle in a periodic potential, one can consider the energy $E_0(\tilde M  +M_0)$ as a Hamiltonian   \cite{LLstPh2}, 
which yields the Hamilton equations 
\be
{d \tilde M\over dt}=- {\partial E_0\over \partial \varphi}=0,
   \ee{Mt}
\be
\omega={d \varphi\over dt}={\partial E_0(\tilde M  +M_0)\over \partial \tilde M}.
            \ee{omega}
While  the canonical quasimoment $\tilde M$ does not vary in time,  the kinetic quasimoment $\tilde m=\tilde M  +M_0$ depends on time:
\be 
{d\tilde m\over dt}={d\tilde M\over dt}  +\dot M_0=\dot M_0.
      \ee{tlM}
In general, these equations must be operator equations for the conjugate operators of the canonical quasimoment   $\tilde M$ and of the angle $\varphi$  \cite{LLstPh2}. But if the torque is weak, one can use the adiabatic approximation with $M_0$ being a slowly varying adiabatic parameter. This allows to assume that at any moment the state does not differ essentially from the eigenstate of  the canonical quasimoment with the eigenvalue $\tilde M$  at fixed $M_0$. Then Eqs.~(\ref{Mt}--\ref{tlM}) can be treated as classical equations.

 In the solid body theory the classical treatment of Eqs.~(\ref{Mt}) and  (\ref{omega}) is sometimes justified by considering them as written for  semiclassical wave packets \cite{ziman}. Since for the quantum rotator the concept of wave packets is problematic (see Sec.~\ref{WP}), it is important that for this  justification we used the adiabatic principle, but not the concept of wave packets.

The function $E_0(\tilde m)$ is determined by the solution of the Schr\"odinger equation in Mathieu functions. Close to the bottom of the band
\be
 E_0(\tilde M  +M_0)={(\tilde M  +M_0)^2\over 2 J^*},~~\omega={\tilde M  +M_0\over J^*},
     \ee{}
where 
\be
J^* = \left[{\partial ^2E_0(\tilde M  +M_0)\over \partial \tilde M^2}\right]^{-1}
    \ee{}
is  the effective moment of inertia, an analog of the effective mass in the Bloch theory for solids.

In the weak binding limit $G \ll \hbar^2/J$ the energy in the Brillouin zone $-\hbar/2 < \tilde m < \hbar/2$ is $E_0={\tilde m^2\over 2J}$ excepting the close vicinity to the zone borders.
The effective moment of inertia $J^*$ does not differ from the bare moment of inertia $J$. The dependence of the energy on the external moment $M_0$  in the weak binding limit is shown for two Bloch bands at quantized values of $\tilde M=s\hbar$ in Fig.~\ref{f1}(b).

In the strong binding limit $G \gg \hbar^2/J$ there is the narrow band
 \be
   E_0=\Delta \left(1-\cos {2\pi \tilde m\over \hbar}\right), 
      \ee{}
where $\Delta $ is the band half-width, which goes to zero at $GJ/\hbar^2\to \infty$. The effective  moment of inertia is
\be
J^*={\hbar^2 \over 4\pi^2\Delta}.
   \ee{}

The band energy has extrema at $M_0=s\hbar$, where the phase angular velocity $\omega$ vanishes. Thus, at zero external moment $M_0$, i.e., in the absence of any connection with the environment, either at the present moment, or in the past, the monotonic phase rotation is impossible. This follows  from the analysis of the quantum pendulum  
\cite{Cond} and of the quantum rotator in a constant electric field \cite{silver} ignoring the connection  with the environment. 
 
 Impossibility of monotonic phase rotation without any connection with the environment can be explained by the following arguments. In  an axisymmetric rotator, i.e., without an external constant field, there are two degenerate eigenstates  of fixed energy, which are either two states $e^{\pm i M\varphi/\hbar}$ with rotating phase, or states $\cos {M\varphi\over \hbar}$ and  $\sin {M\varphi\over \hbar}$  with  vanishing average angular velocity $\langle\dot \varphi\rangle$. But in a coherent superposition of  degenerate states $\cos {M\varphi\over \hbar}$ and  $\sin {M\varphi\over \hbar}$ the  nonzero angular velocity $\langle\dot \varphi\rangle$ is possible. However, whatever weak phase dependent potential (even a single impurity)  breaks the axial symmetry  and lifts degeneracy of states $\cos {M\varphi\over \hbar}$ and  $\sin {M\varphi\over \hbar}$. Then the superposition of two states  is not an eigenstate of the energy operator. In any eigenstate of the energy phase rotation is impossible.

Let us apply  a constant weak torque $N$ to the rotator.  The external moment $M_0=Nt$  is proportional to time, and the time  can be excluded from Eqs.~(\ref{omega}) and (\ref{tlM}). This yields the equation 
\be
N{d \varphi\over d\tilde m}={\partial E_0(\tilde m)\over \partial \tilde m}.
            \ee{Tm}
The equation describes Bloch oscillations with the  time period
 \be
 T={\hbar \over N }. 
       \ee{t}
While in the absence of the external field ($G=0$), the torque produces rotation with acceleration as in the classical theory (Sec.~\ref{per}), in the presence of the periodic external field
the group velocity  $\partial E_0(\tilde m)/\partial \tilde m$ is a periodic function of $\tilde m$, and  the phase  angular velocity $\omega$ performs periodic Bloch oscillations.  The angular velocity   vanishes after averaging over the time  and the amplitude of the phase oscillation is  determined by the  width $\Delta E_0=E_{0\mbox{max}}-E_{0\mbox{min}}$ of the Bloch band:  
\be
\Delta \varphi ={ \Delta E_0\over N }.
     \ee{}
Thus, any finite torque makes  a monotonic rotation impossible. However, in the limit $N \to 0$ when the period  $ T$ becomes much longer than the time of observation, one cannot discern the Bloch oscillation from monotonic rotation.

Next we  phenomenologically introduce dissipation. The environment not  only pumps the moment into the rotator, but also provides a friction torque proportional to the phase angular velocity:
\be
\dot M_0= N -f  \omega = N -f {\partial E_0(\tilde m)\over \partial \tilde m}.
        \ee{dis}
Here $f$ is the friction coefficient. This equation has a solution with constant   $M_0$ and  the phase angular velocity 
\be
\omega={N\over f}.
     \ee{Tf}
Note that  no moment is pumped to the rotator since the moment $M_0$ does not vary in time. The external torque $ N$ is balanced by the friction torque $f\omega$, i.e., the pumped moment  is returned back to the environment.

Rotation of the particle of charge $q$ with the angular velocity $\omega$ produces the current $j=q\omega/2\pi$. At the same time,  the torque is connected with the electric field $E$ [see \eq{cor}]. Thus, \eq{Tf} is in fact the Ohm law $j=El /R_r$, where 
\be
R_r ={4\pi^2 f\over q^2}
     \ee{Rf}
is the resistance to the circular current $j$ around the 1D ring of the quantum rotator.

\begin{figure}[!t]
\includegraphics[width=0.4 \textwidth]{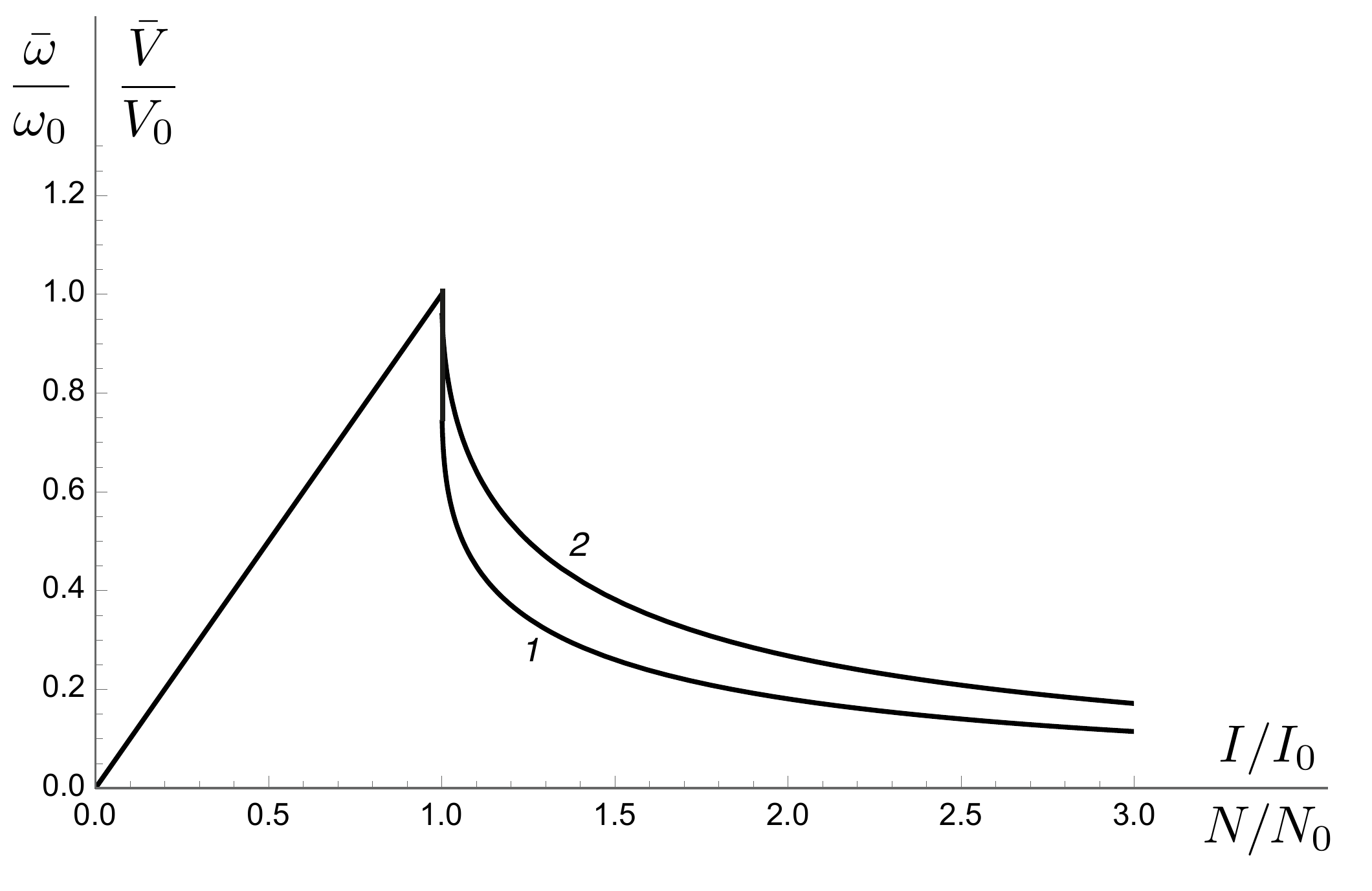} 
 \caption{Plot the average angular velocity $\bar \omega$ vs. torque $N$ for the quantum rotator and plot the average voltage $\bar V$ vs. current $I$ for the JJ in dimensionless variables.  Curve 1:  Weak binding limit, $\omega_0=2\hbar/J$, $N_0=\hbar f/2 J$, $V_0=e/C$, $I_0=e/RC$. Curve 2: Strong binding limit, $\omega_0=\hbar/2\pi J^*$, $N_0=\hbar f/2\pi  J^*$, $V_0=e/\pi C^*$, $I_0=e/\pi RC$.       \label{f2}}
 \end{figure}

The derivative $\partial E_0(\tilde m)/ \partial \tilde m$ of the periodic function has a maximum which determines the  maximum of the phase angular velocity $\omega_m$. When the torque $N$ becomes larger than $f  \omega_m$ the steady phase rotation is  impossible and the Bloch oscillation starts. In the presence of the dissipation Eqs.~(\ref{Tm}) and (\ref{t}) become
\be
{d \varphi\over d\tilde m}=\frac{{\partial E_0(\tilde m)\over \partial \tilde m}}{N-f\ {\partial E_0(\tilde m)\over \partial \tilde m}},
            \ee{Tmf}
 \be
 T=\int\limits_{-\hbar/2}^{\hbar/2}{d\tilde m \over N-f {\partial E_0(\tilde m)\over \partial \tilde m }}. 
       \ee{tf}
Now the phase not only oscillates but also rotates with the average angular velocity
\be
\bar \omega ={1\over T} \int\limits_{-\hbar/2}^{\hbar/2}{{\partial E_0(\tilde m)\over \partial \tilde m } \over N-f {\partial E_0(\tilde m)\over \partial \tilde m }}d\tilde m={N\over f}-{\hbar \over fT}.
   \ee{omf}
At large $N$ the average angular velocity decreases as $1/N$. 

In the weak binding limit $G \ll \hbar^2/J$ 
\be
T= {J\over f} \ln \frac{1+{\hbar f\over 2N J}}{1-{\hbar f\over 2N J}} ,~~\bar \omega={N \over  f}- {\hbar  \over J}\ln^{-1} \frac{1+{\hbar f\over 2N J}}{1-{\hbar f\over 2N J}} .
     \ee{WC}
 In the strong binding limit $G \gg \hbar^2/J$
\be
T= {J^*\over f}{1 \over \sqrt{{J^{*2}N^2\over \hbar^2 f^2 }- {1\over 4\pi^2}}},~~\bar \omega ={  N\over f }-{\hbar \over J^*}\sqrt{{J^{*2}N^2\over \hbar^2 f^2 }- {1\over 4\pi^2}} .
     \ee{WCs}
The dependences of the average angular velocity $\bar\omega$ on the torque $N$ in the weak and the strong limit are shown in the dimensionless variables  in Fig.~\ref{f2}. 

The gauge transformation \eq{gauge} in the variables ``phase--moment'' is 
\be 
\psi =\psi'e^{-M_0\varphi/\hbar }.
      \ee{gaug}
It transforms the Hamiltonian \eq{Hms} and the  Schr\"odinger equation \eq{Sh} to 
\be
H={\hbar^2\over 2 J}\left|{\partial \psi' \over \partial \varphi}\right|^2+[G (1-\cos \varphi )- \dot M_0 \varphi]|\psi'|^2,
            \ee{Hwb}
\be
i\hbar {\partial \psi' \over \partial t}=-{\hbar^2\over 2J}{\partial^2 \psi' \over \partial \varphi^2}+[G (1-\cos \varphi )- \dot M_0 \varphi] \psi',
   \ee{Sht}
which are not periodic in $\varphi$. The  Bloch  wave function after the transformation is also non-periodic:
\bem
 \psi'(\varphi)=u(\tilde M+M_0,\varphi)e^{i \left[(\tilde M+M_0) \varphi- \int^{t}E_0( t')dt'\right]/\hbar}
 \nonumber \\
=u(\tilde M+M_0,\varphi)e^{i\left[\tilde M\varphi-\int^{t}E'_0 (t')dt' \right]/\hbar}
     \eem{M0}
where
\be
E'_0=E_0(\tilde M+M_0) -\dot M_0 \varphi=E_0(\tilde M+M_0) -N\varphi
     \ee{}
is  the energy in the non-periodic gauge.
Further discussion of the non-periodic gauge does not differ essentially from the discussion of this gauge for the axisymmetric rotator (Sec.~\ref{per}). The wave function is not periodic because in the non-periodic gauge the  energy with the washboard potential  is non-periodic.

 \begin{figure}[!t]
\includegraphics[width=0.4 \textwidth]{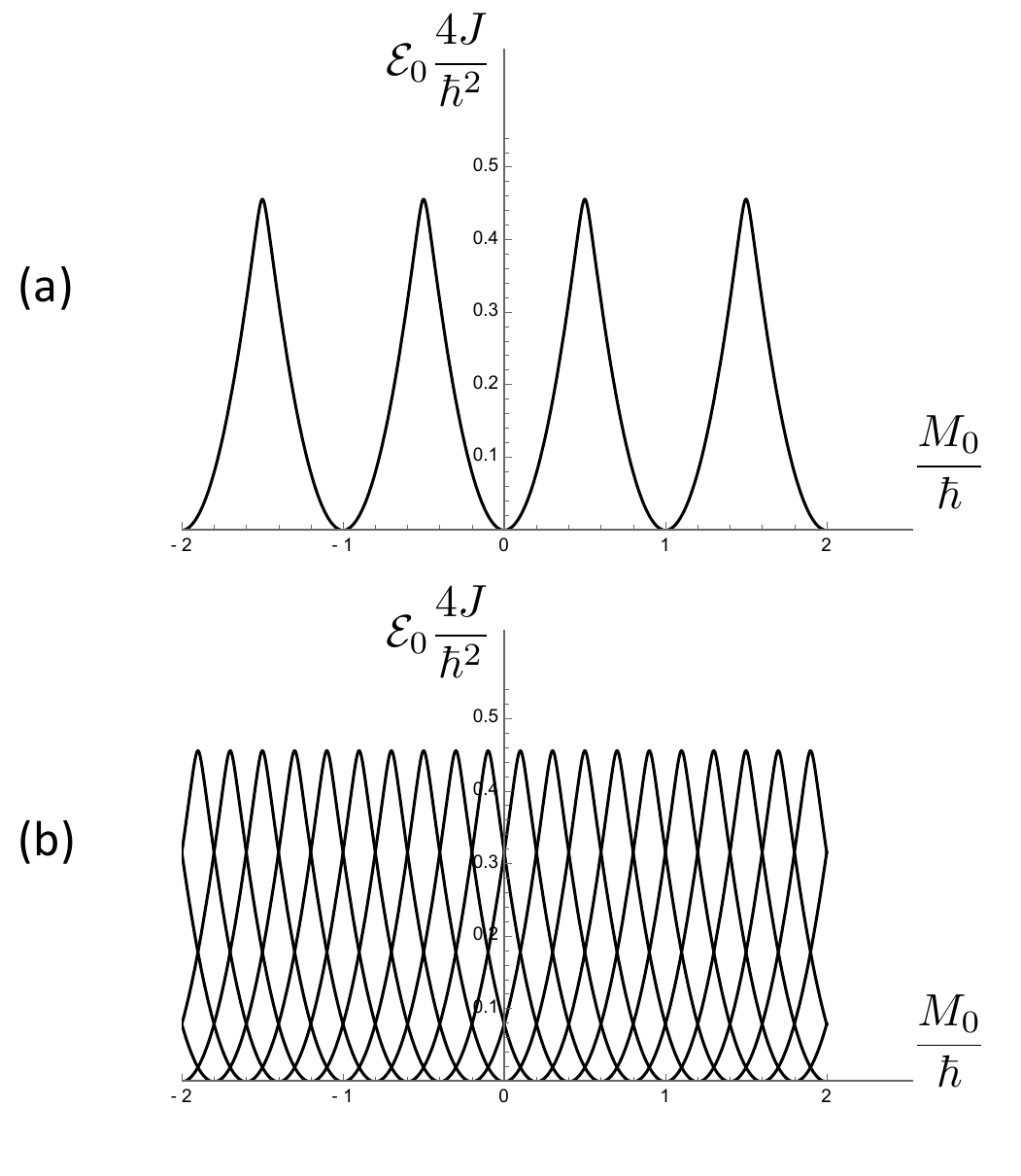} 
 \caption{The band energy ${\cal E}_0$ vs. the external moment $M_0$. (a) The canonical quasimoment $\tilde M =s\hbar$ is quantized. (b) The canonical quasimoment $\tilde M $ is not quantized. There is a continuous manifold of curves for various values of $\tilde M$.       \label{f1.1}}
 \end{figure}

 Let us address now the case of the non-quantized canonical quasimoment $\tilde M$ (particle in a periodic potential in the infinite 1D space). Figure \ref{f1.1} compares the dependence of the energy ${\cal E}_0$ in  the lowest Bloch band on the external moment $M_0$ for the quantized [Fig.~\ref{f1.1}(a)] and the non-quantized  [Fig.~\ref{f1.1}(b)] canonical quasimoment $\tilde M$.  Instead of a single curve for quantized  $\tilde M$ one has a continuous manifold of curves. Figure~\ref{f1.1}(b), however, shows only a discrete manifold of curves in order to demonstrate that the curves are obtained from the single curve in Fig.~\ref{f1.1}(a) by a shift without deformation.

 Despite this difference between quantized and non-quantized $\tilde M$, for the regimes discussed above (monotonous phase rotation and Bloch oscillation)  the effect of quantization is practically absent.  The dynamics of these regimes is governed by the kinetic quasimoment $\tilde m =\tilde M+M_0$. In the absence of quantization the division of  $\tilde m$ into  quantized $\tilde M$ and non-quantized $M_0$ is meaningless since the both are not quantized. If in Fig.~\ref{f1.1}(b) one plots the energy as a function of $\tilde m$ instead of $M_0$ this yields the same single curve as in  Fig.~\ref{f1.1}(a).

Summarizing, the slow dynamics of the particle moving in the 1D ring of the rotator with indistinguishable states with the phases $\varphi$ and $\varphi+2\pi$ does not differ from the dynamics of the particle moving in the infinite 1D space with the periodic potential, when the phases $\varphi$ and $\varphi+2\pi$ correspond to different states.
This is because only one Bloch state participates  in the adiabatic processes. During its tuning by the external torque there are neither  intraband,  nor interband
transitions between Bloch states.

\section{JJ and DQPT}

There is one to one correspondence (ideal mapping) between the quantum rotator in the constant external field and the single JJ. The correspondence between variables of two systems is shown in Table 1. 

\begin{table}[ht]%\centering
\begin{tabular}{|l|l|} \hline  {\bf Rotator}~~~~~~~~~~~~~~~~ & {\bf JJ}
 \\ %\hline 
  &   \\ \hline
 Canonical moment $M$ & Canonical charge $Q \to {2e \over \hbar} M$ \\ \hline
  External moment $M_0$ &  External charge $Q_0 \to {2e \over \hbar}M_0$ \\ \hline 
 Torque $N$ &  Electrical current $I\to {2e \over \hbar}N $ \\ \hline
 Rotation angle $\varphi$ & Quantum mechanical phase $\varphi$ \\ \hline
Angular velocity $\omega=\dot  \varphi$ & Voltage  $V= {\hbar\over 2e}\dot \varphi$ \\ \hline
Moment of inertia $J$ & Capacitance  $C\to {4e ^2\over \hbar^2}J$ \\ \hline
Friction coefficient $f$ & Conductance $1/R\to {4e ^2\over \hbar^2 }f$  \\ \hline
 \end{tabular}
\caption{Correspondence of variables of the rotator and the JJ} \end{table}

As well as in the case of quantum rotator, the theory of the JJ can use either  the periodic gauge, in which the Hamiltonian and the wave function are periodic in $\varphi$, or  the non-periodic gauge, in which both the Hamiltonian and the wave function  are not periodic in $\varphi$. 

Translating the periodic  Hamiltonian \eq{Hms} and the  Schr\"odinger equation \eq{Sh} for the rotator to the JJ one obtains
\be
H={1\over 2 C}\left|-i 2e{\partial \psi \over \partial \varphi} +Q_0\psi\right|^2+E_J (1-\cos \varphi ) |\psi|^2,
            \ee{HmQ}
\be
i\hbar {\partial \psi \over \partial t}=-{1\over 2C}\left(2e{\partial \over \partial \varphi}+iQ_0\right)^2\psi +E_J (1-\cos \varphi ) \psi.
   \ee{ShQ}
Here $Q_0$ is the external charge, while the canonical charge is an operator
\be
\hat Q =-2ie{\partial \over \partial \varphi}.
    \ee{}
The gauge transformation analogous to \eq{gauge},
\be
\psi=\psi' e^{-Q_0 \varphi /2e},
     \ee{}
yields the   Hamiltonian and the  Schr\"odinger equation  in the non-periodic gauge. The wave function  is periodic in $\varphi$ in the periodic gauge, but not periodic in the non-periodic gauge. 

Solutions of the  Schr\"odinger equation \eq{ShQ} are Bloch functions 
\be 
\psi(\varphi,t) =u(\tilde Q+Q_0)e^{i(\tilde Q\varphi/2e -E_0t/\hbar)},
      \ee{}
where  $\tilde Q$ is the canonical quasicharge, which is quantized in the JJ, and $u(\tilde Q+Q_0)$ is a periodic function of the kinetic quasicharge $\tilde q= \tilde Q+Q_0$ with the period $2e$.

The further analysis is similar to that for the quantum rotator (Sec.~\ref{rcf}). The analog of  \eq{dis} is Kirchhoff's law
\be
\dot {\tilde q}=\dot  Q_0 =I - {V\over R},
   \ee{}
where 
 \be
V={\hbar \over 2e}{d\varphi\over dt} 
     \ee{}
is the voltage drop across the JJ. The ohmic resistance 
\be
R= {R_sR_{qp}\over R_s+R_{qp}},
   \ee{}
is determined by the resistance $R_{qp}$ of the normal channel in the JJ and by the resistance $R_s$ of the external shunt parallel to the JJ.

At small current $I$ $\dot Q_0=0$, and the phase rotates with the constant angular velocity, i.e., at the constant voltage $V=IR$. This means that the whole current goes through the ohmic channel, and the JJ is an insulator. The insulating state is possible as far as the voltage $V$ dies not exceeds the voltage $V_0$ equal to the maximum of the derivative ${\partial E_0/ \partial Q_0}$ in the Bloch band. In the limits of weak and strong binding
\be
V_0=\left\{\begin{array}{ccc}  {e\over C} & &   E_J\ll {e^2\over C} \\ {e\over\pi C^*}&  & E_J\gg {e^2\over C}   \end{array}\right.  .
   \ee{}
The voltage $V_0$ is  the electric breakdown voltage of the insulator \cite{CommMurani}. At $V>V_0$  the steady rotation of the phase is impossible. Instead the Bloch oscillation regime takes place accompanied by a slow drift of the phase. The JJ becomes a conductor.  The $VI$ curve of the JJ  is described by the same plot as the plot ``angular velocity vs. torque'' for the quantum rotator shown in Fig.~\ref{f2}. The $VI$ curves in Fig.~\ref{f2} were calculated for the JJ by \citet{widom2} and  by Averin, Likharev, and Zorin \cite{AverLikh,Likh}. \citet{widom2} called the voltage maximum on this curve current-voltage anomaly. 
\citet{Schon} called it Bloch nose. The corresponding maximum on the curve ``resistance--current'' was called the Coulomb blockade bump \cite{SI,Penttila2001,CommMurani}.

While \citet{widom2}  calculated the $VI$ curve using the analogy with the pendulum, i.e., assuming that the states with the phases $\varphi$ and $\varphi+2\pi$ are  identical, Averin, Likharev, and Zorin \cite{AverLikh,Likh} assumed that they are not identical as in the case of  a particle in a periodic potential. Nevertheless, the both groups obtained the same $VI$ curves in agreement with our conclusion in the end of Sec.~\ref{rcf}. 

The possibility of monotonic phase rotation in the Bloch band theory is due to quantum tunneling between neighboring wells of the periodic potential. Dissipation can suppress quantum tunneling \cite{CL}. Then the particle (virtual particle in the JJ case) becomes localized in one of the wells of the periodic potential. This is the superconducting state of the JJ. The transition from the superconducting to the insulating state is the DQPT. 

The DQPT is a joint effect of Coulomb interaction, dissipation, and quantum mechanics. The Coulomb blockade of Cooper pairs makes the JJ an insulator at small bias. However, it is effective only if the Coulomb energy $E_C= e^2/C$ exceeds the quantum-mechanical uncertainty $\hbar /\tau$ \cite{Tin}, where $\tau=RC$ is the time of the charge relaxation in the circuit.  According to the condition $E_C \sim \hbar/\tau$, the DQPT is expected at the resistance $R$ of the order of the quantum resistance  $R_q=h/4e^2$. This qualitative estimation \cite{Tin,CommMurani} agrees with the more detailed and accurate theory predicting the DQPT  exactly at $R = R_q$  \cite{Schmid,Bulg}.   
 
A similar DQPT must exist in the quantum rotator. The analog of the Coulomb energy $e^2/C$  is the energy 
$\hbar^2/ J$ necessary for a transfer of the moment quantum $\hbar$ to the rotator. The time $\tau=J/f$ is the decay time for the moment in the rotator with friction. Thus, the critical friction coefficient is of the order of $f \sim \hbar$. \Eq{Rf} yields the relation between the friction coefficient $f$ and the ohmic resistance $R_r$ for the current produced by the particle of charge $q$ rotating in the rotator. According to this relation, the condition  $f \sim \hbar$ for the DQPT in the rotator is identical to the condition $R_r\sim R_q $. Now $R_q \sim \hbar /q^2$ is the quantum resistance for the charge $q$.

However, there is a difference in the role of the resistance $R$ in the JJ and of the resistance $R_r$ in the quantum rotator. In the quantum rotator the monotonic phase rotation is possible at small $R_r<R_q$, while in the JJ the phase rotates at large $R>R_q$. In the regime of the phase rotation the JJ is an insulator, while the quantum rotator is a conductor. In the regime of the localized phase the JJ is a superconductor, while the quantum rotator is an insulator. In the both cases the insulating state takes place at  resistances larger than the quantum resistance.

One can estimate the resistance of the quantum rotator using the Drude formula for the conductivity $l/R_r$, which for the 1D system in the weak-binding limit yields
\be
{l\over R_r}\sim  {q^2\tau \over m_0}={q^2l_0 \over \hbar k l},
    \ee{}
where $l_0$ is the mean-free path of the particle and $\tau =l_0/v=m_0l_0/\hbar k$ is the relaxation time at elastic scattering by impurities. Since the wave number $k$ of the particle is on the order of the inverse space period $l$, the phase transition condition 
\be
{R_q\over R_r}\sim {l_0 \over l^2 k} \sim kl_0 \sim 1
   \ee{}
becomes the Ioffe--Regel condition for the metal--insulator transition \cite{Mott}.

Although the pioneer theoretical investigations \cite{Schmid,Bulg} predicted the DQPT at the line $R=R_q$ independently from the ratio $E_J/ E_C$, later it became clear that in the real experiment it is impossible to detect the transition at this line at $E_J\gg E_C$.  One reason is an inevitable non-zero temperature of the experiment  \cite{Zaik}. But even at strictly zero temperature the insulator state is not observable at $E_J\gg E_C$ because  either  the observation time is short compared with the average time interval between tunneling events, which are phase slips destroying superconductivity \cite{Schon}, or the accuracy of the voltage measurement is not sufficient for detection of the phase rotation since the voltage error bar exceeds the electric breakdown voltage $V_0$ \cite{SI,Penttila2001}. 

\begin{figure}[t]
\includegraphics[width=.45\textwidth]{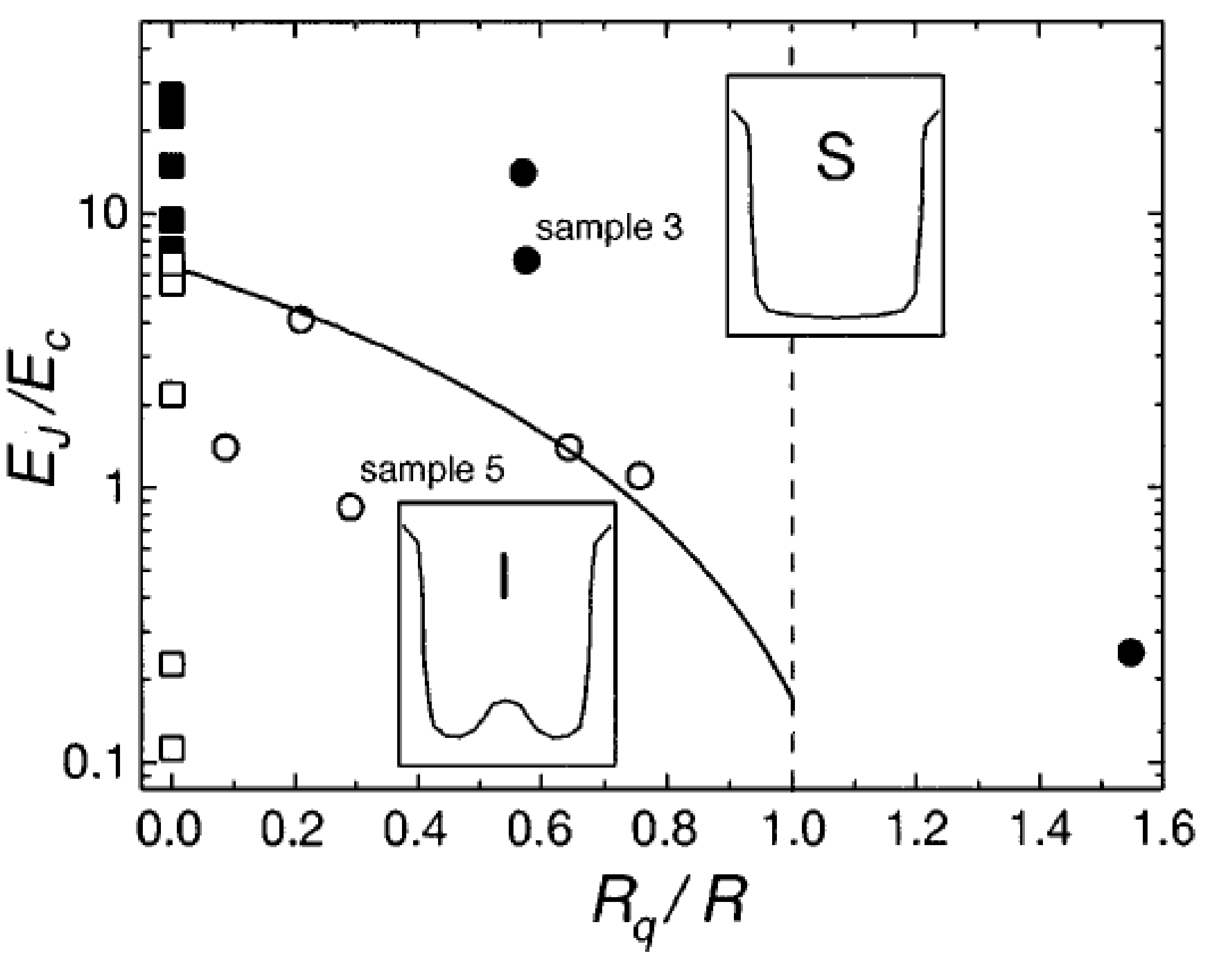}
 \caption{The phase diagram of the JJ \cite{SI}. The dashed vertical line shows the DQPT of  \citet{Schmid} and \citet{Bulg}. Due to  voltage measurement error bar the experimental detection of the DQPT is expected at the solid line  \cite{SI}.
 Black and open symbols show observations of the superconductorlike ($S$) and the insulatorlike ($I$) $RI$ curves (see insets), respectively. Squares shows  results of experimental observations of unshunted junctions when the resistance $R$ is equal to a very large quasiparticle resistance $R_{qp}$ of the junction itself. \footnote{ In the caption to Fig.~3 in Ref.~\onlinecite{SI}  it was stated that “unshunted samples (squares) are collected at  $R_q/R=0$”. However, as said in the text of the paper, because of a large shunting quasiparticle resistance $R_{qp}$, $R_q/R$ was never truly equal to 0 in the experiments.}
 \label{PD}}
 \end{figure}

The results of the experimental and theoretical investigation of the phase diagram by Penttil\"a {\em et al.} \cite{SI,Penttila2001}  are shown in Fig.~\ref{PD}. Open and black circles show observations of the superconductorlike and the insulatorlike behavior, respectively. The insulating  (I) state differs from the superconducting (S) state by the presence of the Coulomb blockade bump at dependences of the resistance on the current.
The solid line was determined from the condition that the error bar of voltage measurements is equal to the voltage $V_0$ at which the electric breakdown of the insulating state takes place. It is impossible to detect the insulating state above the solid line.  The observation of  the Coulomb blockade bump below the solid line  is  a smoking gun of the DQPT.

If one increases measurement accuracy (or lowers temperature), the  solid line  moves closer to the Schmid--Bulgadaev line. Thus, the vertical  Schmid--Bulgadaev line is an idealized asymptotic limit, which remains experimentally unattainable in practice for large $E_J/E_C$.  

\section{Discussion and conclusions}

Although the analogy of the JJ with the quantum rotator was pointed out in the past \cite{AzbelP,Rogov,widom2,Loss}, the opinion that the JJ is an analog of  a particle in a periodic potential was more prevalent   \cite{AverLikh,Likh,GefenJ,Zwerg,Apen,SI,Penttila2001,CommMurani,morel,Schon,golub}. Averin, Likharev, and Zorin \cite{AverLikh,Likh} argued that   the states of the JJ with $\varphi$ and $\varphi+2\pi$ are not identical because the states of the environment (electric circuit) for two values of the phase are different. As a result, they concluded that all states in  the Bloch band are possible and used the concept of wave packets at the derivation  of Bloch oscillations.  \citet{Zwerg} and  \citet{morel} explained the distinguishability of the states  with $\varphi$ and $\varphi+2\pi$ (decompactification of the phase) by the effect of dissipation. \citet{Apen} argued that the distinguishability assumption is justified by taking into account fluctuations of the external classical charge, which were not taken into account in the previous calculations of the $VI$ curve in Refs.~\cite{widom2,AverLikh,Likh} and in the present paper. 

 The assumption that the states of the JJ with $\varphi$ and $\varphi+2\pi$ can be different was disputed by \citet{Loss}. We also believe that the indistinguishability of the states with $\varphi$ and $\varphi+2\pi$ is a just requirement for the quantum rotator and the JJ. Although the connection between the JJ and the environment is of utmost importance and must not be ignored, we consider  the wave function of the junction alone, but not the wave function of the junction+environment.  There is an important difference between two questions: (i) whether the states of the JJ with $\varphi$ and $\varphi+2\pi$ can be different or not, and (ii) whether the states of the JJ and the environment are the same before and after a phase slip, which is  a jump $2\pi$ of the phase difference across the JJ. 
 
Averin, Likharev, and Zorin \cite{AverLikh,Likh}  looked for an answer to the second question. Let us illustrate this for a simple SQUID circuit, which  is a superconducting loop interrupted by a JJ. There is a phase $\varphi$ across the JJ and a phase difference $\Phi$ along the rest part of the loop. They must satisfy the boundary condition $\varphi+\Phi =2\pi N$, where $N$ is an integer. Averin, Likharev, and Zorin compared the state with $\varphi$, $\Phi$,  and $N$,  with  the state with $\varphi+2\pi $, $\Phi-2\pi$,  and $N$. These are states before and after a phase slip, which are definitely different. While before the phase slip the state was stationary, after the slip the state is not stationary: Adding $2\pi$ to the Josephson phase does not change the current across the JJ  but  does change the current in the rest part of the circuit and the magnetic flux because the phase difference $\Phi$ is not the same before and after the slip.

The states of the JJ with the phases $\varphi$ and $\varphi+2\pi $ must be compared at fixed state of the environment. One should compare  the state with $\varphi$,  $\Phi$, and $N$ with the state with $\varphi+2\pi $, $\Phi$, and $N+1$. The environment (the rest part of the circuit) remains in the same state because the phase $\Phi$  remains the same, and the current and the magnetic flux do not differ.
 
 Through the whole paper it was assumed that the external moment in the quantum rotator (charge in the JJ) is a well defined classical variable slowly varying in time.  As mentioned above,
 fluctuations of the external charge were considered as a reason for the phase decompactification in the JJ. Indeed, for any {\em given} external charge only {\em one} state in the Bloch band  is possible, but this state is different for different  external charges. Thus, for a broad ensemble of external charges any state in the Bloch band  is possible as in the case of a particle in the infinite 1D state. However, this does not eliminate the difference between the Hilbert space  of all possible Bloch states in the case of a particle in the infinite 1D state and its much smaller discrete subspace of the rotator states with quantized canonical moments (charges). If one has the ensemble of $n$ external moments, the number of states of the  system ``quantum rotator+environment'' is equal to $n$, while  the number of states for ``particle in an infinite 1D space+environment'' is $n\times n_B$, where $n_B$ is the number of all Bloch states in a Bloch band.
  
 This difference is not important for the conventional theory on the insulating state and the DQPT. This is true because the theory deals only with a slow adiabatic tuning of only {\em one} state in the lowest Bloch band. The question whether this state is a single state or other Bloch states  are possible, is irrelevant. This is another example when  according to \citet{Legg} ``it is largely unnecessary  to address the vexed question of whether or not states differing in $\varphi$ value by $2\pi$ should be identified.''
 
 This does not mean that difference between dynamics of a particle in a 1D ring and a particle in the infinite 1D space is not important in general, at any experimental conditions. 
 \citet{Loss} made calculations of the resonant tunneling between quantum levels in the neighboring potential wells of the washboard potential in the JJ.  In the Bloch theory this is the Zener interband tunneling. The results depended on the choice of the initial state. The latter  was either periodic in the extended phase as must be for a particle in a 1D ring, or was confined in the interval $2\pi$ that is possible only for a particle in the infinite 1D space. This calculation has not yet resolved the dilemma ``compact vs. extended phase''. \citet{Loss}  solved the Schr\"odinger equation \eq{ShQ} at the constant current $I= \dot Q_0$, while the proponents of the extended phase assumption for the JJ connected the decompactification with fluctuations of the external charge $Q_0$, which were not taken into account in their calculation. Thus,  the answer to the question whether and how the difference between motion of a particle in a 1D ring and motion of a particle in the infinite 1D space with a periodic potential can affect observable phenomena  is still lacking.

Arguing that the conventional theory failed and the insulating state is impossible in the JJ, \citet{Sacl}  referred to the existence of supercurrent in the Cooper pair box, which is identical to a JJ in the limit $R\to \infty$. Indeed, in Sec.~\ref{rcf}  we demonstrated that without dissipation the phase (angle) of the quantum rotator is localized and  performs Bloch oscillations around the localization point. Thus, a supercurrent flows across the JJ. However, from the same subsection it is clear that {\em any} dissipation whatever large $R$ be,  delocalizes the phase and the JJ becomes an insulator. One can see in Fig.~\ref {f1.1} that at $R \to \infty $ the voltage vanishes at currents $I\gg e/RC^*$. Although the insulating state is possible at any $R$, the electric breakdown  of the insulator occurs at the current  $I\sim e/RC^*$, which vanishes at $R \to \infty $. This makes observation of the DQPT impossible in this limit, but it is not an argument against its existence.

Summarizing, our answers to three questions formulated in the introduction are following:
\begin{enumerate}
\item
The states with the phases $\varphi$ and $\varphi+2\pi$ are indistinguishable in the JJ for the fixed state of the environment. In this aspect the JJ is an analog of the quantum rotator, but not of a particle in a periodic potential.
\item
The assumption that in the quantum rotator and in the JJ the states  with the phases $\varphi$ and $\varphi+2\pi$ are indistinguishable, does not mean that the wave functions must be also periodic in all gauges. The wave functions can be  periodic only in the periodic gauge where the Hamiltonian is periodic in $\varphi$. 
\item
The conventional  theory of the DQPT is valid independently from whether the states with the phases $\varphi$ and $\varphi+2\pi$ are  distinguishable, or not.
\end{enumerate}

The present analysis and its conclusions referred only to the case of a single particle, which excludes  interaction between particles. The case of many interacting particles in a quantum rotator requires another approach to the dilemma ``compact vs. extended  phase'' \cite{averin}.

The close analogy between the regime of phase rotation in the quantum  rotator and the phase rotation in the insulating state of  the JJ allows to  expect the DQPT also in the quantum rotator. This can be checked by experimental investigations of persistent currents in 1D normal rings put in a constant electric field.

\begin{acknowledgments}
I thank Dmitri Averin, Pertti  Hakonen, and Andrei Zaikin for discussions and useful comments.
\end{acknowledgments}

%\bibliography{bibBook,nano}

\begin{thebibliography}{41}%
\makeatletter
\providecommand \@ifxundefined [1]{%
 \@ifx{#1\undefined}
}%
\providecommand \@ifnum [1]{%
 \ifnum #1\expandafter \@firstoftwo
 \else \expandafter \@secondoftwo
 \fi
}%
\providecommand \@ifx [1]{%
 \ifx #1\expandafter \@firstoftwo
 \else \expandafter \@secondoftwo
 \fi
}%
\providecommand \natexlab [1]{#1}%
\providecommand \enquote  [1]{``#1''}%
\providecommand \bibnamefont  [1]{#1}%
\providecommand \bibfnamefont [1]{#1}%
\providecommand \citenamefont [1]{#1}%
\providecommand \href@noop [0]{\@secondoftwo}%
\providecommand \href [0]{\begingroup \@sanitize@url \@href}%
\providecommand \@href[1]{\@@startlink{#1}\@@href}%
\providecommand \@@href[1]{\endgroup#1\@@endlink}%
\providecommand \@sanitize@url [0]{\catcode `\\12\catcode `\$12\catcode
  `\&12\catcode `\#12\catcode `\^12\catcode `\_12\catcode `\%12\relax}%
\providecommand \@@startlink[1]{}%
\providecommand \@@endlink[0]{}%
\providecommand \url  [0]{\begingroup\@sanitize@url \@url }%
\providecommand \@url [1]{\endgroup\@href {#1}{\urlprefix }}%
\providecommand \urlprefix  [0]{URL }%
\providecommand \Eprint [0]{\href }%
\providecommand \doibase [0]{https://doi.org/}%
\providecommand \selectlanguage [0]{\@gobble}%
\providecommand \bibinfo  [0]{\@secondoftwo}%
\providecommand \bibfield  [0]{\@secondoftwo}%
\providecommand \translation [1]{[#1]}%
\providecommand \BibitemOpen [0]{}%
\providecommand \bibitemStop [0]{}%
\providecommand \bibitemNoStop [0]{.\EOS\space}%
\providecommand \EOS [0]{\spacefactor3000\relax}%
\providecommand \BibitemShut  [1]{\csname bibitem#1\endcsname}%
\let\auto@bib@innerbib\@empty
%</preamble>
\bibitem [{\citenamefont {Azbel}\ and\ \citenamefont {{Per
  Bak}}(1984)}]{AzbelP}%
  \BibitemOpen
  \bibfield  {author} {\bibinfo {author} {\bibfnamefont {M.~Y.}\ \bibnamefont
  {Azbel}}\ and\ \bibinfo {author} {\bibnamefont {{Per Bak}}},\ }\bibfield
  {title} {\bibinfo {title} {Analytical results on the periodically driven
  damped pendulum. {A}pplication to sliding charge-density waves and
  {J}osephson junctions},\ }\href@noop {} {\bibfield  {journal} {\bibinfo
  {journal} {Phys. Rev. B}\ }\textbf {\bibinfo {volume} {30}},\ \bibinfo
  {pages} {3722} (\bibinfo {year} {1984})}\BibitemShut {NoStop}%
\bibitem [{\citenamefont {B{\"u}ttiker}\ \emph {et~al.}(1984)\citenamefont
  {B{\"u}ttiker}, \citenamefont {Imry},\ and\ \citenamefont {Azbel}}]{AzbelN}%
  \BibitemOpen
  \bibfield  {author} {\bibinfo {author} {\bibfnamefont {M.}~\bibnamefont
  {B{\"u}ttiker}}, \bibinfo {author} {\bibfnamefont {Y.}~\bibnamefont {Imry}},\
  and\ \bibinfo {author} {\bibfnamefont {M.~Y.}\ \bibnamefont {Azbel}},\
  }\bibfield  {title} {\bibinfo {title} {Quantum oscillations in
  one-dimensional normal-metal rings},\ }\href@noop {} {\bibfield  {journal}
  {\bibinfo  {journal} {Phys. Rev. A}\ }\textbf {\bibinfo {volume} {30}},\
  \bibinfo {pages} {1982 } (\bibinfo {year} {1984})}\BibitemShut {NoStop}%
\bibitem [{\citenamefont {Gefen}\ \emph {et~al.}(1984)\citenamefont {Gefen},
  \citenamefont {Imry},\ and\ \citenamefont {Azbel}}]{Azbel}%
  \BibitemOpen
  \bibfield  {author} {\bibinfo {author} {\bibfnamefont {Y.}~\bibnamefont
  {Gefen}}, \bibinfo {author} {\bibfnamefont {Y.}~\bibnamefont {Imry}},\ and\
  \bibinfo {author} {\bibfnamefont {M.~Y.}\ \bibnamefont {Azbel}},\ }\bibfield
  {title} {\bibinfo {title} {Quantum oscillations and the {A}haronov-{B}ohm
  effect for parallel resistors},\ }\href
  {https://doi.org/10.1103/PhysRevLett.52.129} {\bibfield  {journal} {\bibinfo
  {journal} {Phys. Rev. Lett.}\ }\textbf {\bibinfo {volume} {52}},\ \bibinfo
  {pages} {129} (\bibinfo {year} {1984})}\BibitemShut {NoStop}%
\bibitem [{\citenamefont {Dirac}(1927)}]{dirac}%
  \BibitemOpen
  \bibfield  {author} {\bibinfo {author} {\bibfnamefont {P.~A.~M.}\
  \bibnamefont {Dirac}},\ }\href@noop {} {\bibfield  {journal} {\bibinfo
  {journal} {Proc. Roy. Soc. (London) Ser. A}\ }\textbf {\bibinfo {volume}
  {114}},\ \bibinfo {pages} {243} (\bibinfo {year} {1927})}\BibitemShut
  {NoStop}%
\bibitem [{\citenamefont {Judge}(1964)}]{judge}%
  \BibitemOpen
  \bibfield  {author} {\bibinfo {author} {\bibfnamefont {D.}~\bibnamefont
  {Judge}},\ }\bibfield  {title} {\bibinfo {title} {On the uncertainty relation
  for angle variables},\ }\href@noop {} {\bibfield  {journal} {\bibinfo
  {journal} {Nuovo Cimento}\ }\textbf {\bibinfo {volume} {31}},\ \bibinfo
  {pages} {332} (\bibinfo {year} {1964})}\BibitemShut {NoStop}%
\bibitem [{\citenamefont {Carruthers}\ and\ \citenamefont
  {Nieto}(1968)}]{PhasRev}%
  \BibitemOpen
  \bibfield  {author} {\bibinfo {author} {\bibfnamefont {P.}~\bibnamefont
  {Carruthers}}\ and\ \bibinfo {author} {\bibfnamefont {M.}~\bibnamefont
  {Nieto}},\ }\bibfield  {title} {\bibinfo {title} {Phase and angle variables
  in quantum mechanics},\ }\href@noop {} {\bibfield  {journal} {\bibinfo
  {journal} {Rev. Mod. Phys.}\ }\textbf {\bibinfo {volume} {40}},\ \bibinfo
  {pages} {411} (\bibinfo {year} {1968})}\BibitemShut {NoStop}%
\bibitem [{\citenamefont {Barnett}\ and\ \citenamefont {Pegg}(1990)}]{Pegg}%
  \BibitemOpen
  \bibfield  {author} {\bibinfo {author} {\bibfnamefont {S.~M.}\ \bibnamefont
  {Barnett}}\ and\ \bibinfo {author} {\bibfnamefont {D.}~\bibnamefont {Pegg}},\
  }\bibfield  {title} {\bibinfo {title} {Quantum theory of rotation angles},\
  }\href@noop {} {\bibfield  {journal} {\bibinfo  {journal} {Phys. Rev. A}\
  }\textbf {\bibinfo {volume} {41}},\ \bibinfo {pages} {3427} (\bibinfo {year}
  {1990})}\BibitemShut {NoStop}%
\bibitem [{\citenamefont {Susskind}\ and\ \citenamefont
  {Glogower}(1964)}]{suss}%
  \BibitemOpen
  \bibfield  {author} {\bibinfo {author} {\bibfnamefont {L.}~\bibnamefont
  {Susskind}}\ and\ \bibinfo {author} {\bibfnamefont {J.}~\bibnamefont
  {Glogower}},\ }\bibfield  {title} {\bibinfo {title} {Quantum mechanical phase
  and time operator},\ }\href@noop {} {\bibfield  {journal} {\bibinfo
  {journal} {Physica}\ }\textbf {\bibinfo {volume} {1}},\ \bibinfo {pages} {49}
  (\bibinfo {year} {1964})}\BibitemShut {NoStop}%
\bibitem [{\citenamefont {Murani}\ \emph {et~al.}(2020)\citenamefont {Murani},
  \citenamefont {Bourlet}, \citenamefont {le~Sueur}, \citenamefont {Portier},
  \citenamefont {Altimiras}, \citenamefont {Esteve}, \citenamefont {Grabert},
  \citenamefont {Stockburger}, \citenamefont {Ankerhold},\ and\ \citenamefont
  {Joyez}}]{Sacl}%
  \BibitemOpen
  \bibfield  {author} {\bibinfo {author} {\bibfnamefont {A.}~\bibnamefont
  {Murani}}, \bibinfo {author} {\bibfnamefont {N.}~\bibnamefont {Bourlet}},
  \bibinfo {author} {\bibfnamefont {H.}~\bibnamefont {le~Sueur}}, \bibinfo
  {author} {\bibfnamefont {F.}~\bibnamefont {Portier}}, \bibinfo {author}
  {\bibfnamefont {C.}~\bibnamefont {Altimiras}}, \bibinfo {author}
  {\bibfnamefont {D.}~\bibnamefont {Esteve}}, \bibinfo {author} {\bibfnamefont
  {H.}~\bibnamefont {Grabert}}, \bibinfo {author} {\bibfnamefont
  {J.}~\bibnamefont {Stockburger}}, \bibinfo {author} {\bibfnamefont
  {J.}~\bibnamefont {Ankerhold}},\ and\ \bibinfo {author} {\bibfnamefont
  {P.}~\bibnamefont {Joyez}},\ }\bibfield  {title} {\bibinfo {title} {Absence
  of a dissipative quantum phase transition in {J}osephson junctions},\ }\href
  {https://doi.org/10.1103/PhysRevX.10.021003} {\bibfield  {journal} {\bibinfo
  {journal} {Phys. Rev. X}\ }\textbf {\bibinfo {volume} {10}},\ \bibinfo
  {pages} {021003} (\bibinfo {year} {2020})}\BibitemShut {NoStop}%
\bibitem [{\citenamefont {Hakonen}\ and\ \citenamefont
  {Sonin}(2021)}]{CommMurani}%
  \BibitemOpen
  \bibfield  {author} {\bibinfo {author} {\bibfnamefont {P.~J.}\ \bibnamefont
  {Hakonen}}\ and\ \bibinfo {author} {\bibfnamefont {E.~B.}\ \bibnamefont
  {Sonin}},\ }\bibfield  {title} {\bibinfo {title} {Comment on ``{A}bsence of a
  dissipative quantum phase transition in josephson junctions''},\ }\href
  {https://doi.org/10.1103/PhysRevX.11.018001} {\bibfield  {journal} {\bibinfo
  {journal} {Phys. Rev. X}\ }\textbf {\bibinfo {volume} {11}},\ \bibinfo
  {pages} {018001} (\bibinfo {year} {2021})}\BibitemShut {NoStop}%
\bibitem [{\citenamefont {Murani}\ \emph {et~al.}(2021)\citenamefont {Murani},
  \citenamefont {Bourlet}, \citenamefont {le~Sueur}, \citenamefont {Portier},
  \citenamefont {Altimiras}, \citenamefont {Esteve}, \citenamefont {Grabert},
  \citenamefont {Stockburger}, \citenamefont {Ankerhold},\ and\ \citenamefont
  {Joyez}}]{ReplMurani}%
  \BibitemOpen
  \bibfield  {author} {\bibinfo {author} {\bibfnamefont {A.}~\bibnamefont
  {Murani}}, \bibinfo {author} {\bibfnamefont {N.}~\bibnamefont {Bourlet}},
  \bibinfo {author} {\bibfnamefont {H.}~\bibnamefont {le~Sueur}}, \bibinfo
  {author} {\bibfnamefont {F.}~\bibnamefont {Portier}}, \bibinfo {author}
  {\bibfnamefont {C.}~\bibnamefont {Altimiras}}, \bibinfo {author}
  {\bibfnamefont {D.}~\bibnamefont {Esteve}}, \bibinfo {author} {\bibfnamefont
  {H.}~\bibnamefont {Grabert}}, \bibinfo {author} {\bibfnamefont
  {J.}~\bibnamefont {Stockburger}}, \bibinfo {author} {\bibfnamefont
  {J.}~\bibnamefont {Ankerhold}},\ and\ \bibinfo {author} {\bibfnamefont
  {P.}~\bibnamefont {Joyez}},\ }\bibfield  {title} {\bibinfo {title} {Reply to
  ``{C}omment on ``{A}bsence of a dissipative quantum phase transition in
  {J}osephson junctions'''},\ }\href
  {https://doi.org/10.1103/PhysRevX.11.018002} {\bibfield  {journal} {\bibinfo
  {journal} {Phys. Rev. X}\ }\textbf {\bibinfo {volume} {11}},\ \bibinfo
  {pages} {018002} (\bibinfo {year} {2021})}\BibitemShut {NoStop}%
\bibitem [{\citenamefont {Schmid}(1983)}]{Schmid}%
  \BibitemOpen
  \bibfield  {author} {\bibinfo {author} {\bibfnamefont {A.}~\bibnamefont
  {Schmid}},\ }\bibfield  {title} {\bibinfo {title} {Diffusion and localization
  in a dissipative quantum system},\ }\href
  {https://doi.org/10.1103/PhysRevLett.51.1506} {\bibfield  {journal} {\bibinfo
   {journal} {Phys. Rev. Lett.}\ }\textbf {\bibinfo {volume} {51}},\ \bibinfo
  {pages} {1506} (\bibinfo {year} {1983})}\BibitemShut {NoStop}%
\bibitem [{\citenamefont {Bulgadaev}(1984)}]{Bulg}%
  \BibitemOpen
  \bibfield  {author} {\bibinfo {author} {\bibfnamefont {S.~A.}\ \bibnamefont
  {Bulgadaev}},\ }\bibfield  {title} {\bibinfo {title} {Phase diagram of a
  dissipative quantum system},\ }\href@noop {} {\bibfield  {journal} {\bibinfo
  {journal} {JETP Lett.}\ }\textbf {\bibinfo {volume} {39}},\ \bibinfo {pages}
  {315} (\bibinfo {year} {1984})}\BibitemShut {NoStop}%
\bibitem [{\citenamefont {Condon}(1928)}]{Cond}%
  \BibitemOpen
  \bibfield  {author} {\bibinfo {author} {\bibfnamefont {E.~U.}\ \bibnamefont
  {Condon}},\ }\bibfield  {title} {\bibinfo {title} {The physical pendulum in
  quantum mechanics},\ }\href {https://doi.org/10.1103/PhysRev.31.891}
  {\bibfield  {journal} {\bibinfo  {journal} {Phys. Rev.}\ }\textbf {\bibinfo
  {volume} {31}},\ \bibinfo {pages} {891} (\bibinfo {year} {1928})}\BibitemShut
  {NoStop}%
\bibitem [{\citenamefont {Silverman}(1981)}]{silver}%
  \BibitemOpen
  \bibfield  {author} {\bibinfo {author} {\bibfnamefont {M.~P.}\ \bibnamefont
  {Silverman}},\ }\bibfield  {title} {\bibinfo {title} {Exact spectrum of the
  two-dimensional rigid rotator in external fields. {I. S}tark effect},\ }\href
  {https://doi.org/10.1103/PhysRevA.24.339} {\bibfield  {journal} {\bibinfo
  {journal} {Phys. Rev. A}\ }\textbf {\bibinfo {volume} {24}},\ \bibinfo
  {pages} {339} (\bibinfo {year} {1981})}\BibitemShut {NoStop}%
\bibitem [{\citenamefont {B{\"u}ttiker}\ \emph {et~al.}(1983)\citenamefont
  {B{\"u}ttiker}, \citenamefont {Imry},\ and\ \citenamefont
  {Landauer}}]{Buttiker}%
  \BibitemOpen
  \bibfield  {author} {\bibinfo {author} {\bibfnamefont {M.}~\bibnamefont
  {B{\"u}ttiker}}, \bibinfo {author} {\bibfnamefont {Y.}~\bibnamefont {Imry}},\
  and\ \bibinfo {author} {\bibfnamefont {R.}~\bibnamefont {Landauer}},\
  }\bibfield  {title} {\bibinfo {title} {Josephson behavior in small normal
  one-dimensional rings},\ }\href
  {https://doi.org/https://doi.org/10.1016/0375-9601(83)90011-7} {\bibfield
  {journal} {\bibinfo  {journal} {Physics Letters A}\ }\textbf {\bibinfo
  {volume} {96}},\ \bibinfo {pages} {365 } (\bibinfo {year}
  {1983})}\BibitemShut {NoStop}%
\bibitem [{\citenamefont {Cheung}\ \emph {et~al.}(1988)\citenamefont {Cheung},
  \citenamefont {Gefen}, \citenamefont {Riedel},\ and\ \citenamefont
  {Shih}}]{Gefen}%
  \BibitemOpen
  \bibfield  {author} {\bibinfo {author} {\bibfnamefont {H.-F.}\ \bibnamefont
  {Cheung}}, \bibinfo {author} {\bibfnamefont {Y.}~\bibnamefont {Gefen}},
  \bibinfo {author} {\bibfnamefont {E.~K.}\ \bibnamefont {Riedel}},\ and\
  \bibinfo {author} {\bibfnamefont {W.-H.}\ \bibnamefont {Shih}},\ }\bibfield
  {title} {\bibinfo {title} {Persistent currents in small one-dimensional metal
  rings},\ }\href {https://doi.org/10.1103/PhysRevB.37.6050} {\bibfield
  {journal} {\bibinfo  {journal} {Phys. Rev. B}\ }\textbf {\bibinfo {volume}
  {37}},\ \bibinfo {pages} {6050} (\bibinfo {year} {1988})}\BibitemShut
  {NoStop}%
\bibitem [{\citenamefont {Loss}\ and\ \citenamefont {Mullen}(1991)}]{LossM}%
  \BibitemOpen
  \bibfield  {author} {\bibinfo {author} {\bibfnamefont {D.}~\bibnamefont
  {Loss}}\ and\ \bibinfo {author} {\bibfnamefont {K.}~\bibnamefont {Mullen}},\
  }\bibfield  {title} {\bibinfo {title} {Effect of dissipation on phase
  periodicity and the quantum dynamics of {J}osephson junctions},\ }\href@noop
  {} {\bibfield  {journal} {\bibinfo  {journal} {Phys. Rev. A}\ }\textbf
  {\bibinfo {volume} {43}},\ \bibinfo {pages} {2129} (\bibinfo {year}
  {1991})}\BibitemShut {NoStop}%
\bibitem [{\citenamefont {Landau}\ and\ \citenamefont {Lifshitz}(1982)}]{LLqu}%
  \BibitemOpen
  \bibfield  {author} {\bibinfo {author} {\bibfnamefont {L.~D.}\ \bibnamefont
  {Landau}}\ and\ \bibinfo {author} {\bibfnamefont {E.~M.}\ \bibnamefont
  {Lifshitz}},\ }\href@noop {} {\emph {\bibinfo {title} {Quantum mechanics}}}\
  (\bibinfo  {publisher} {Pergamon Press},\ \bibinfo {year} {1982})\BibitemShut
  {NoStop}%
\bibitem [{\citenamefont {Davidson}\ and\ \citenamefont
  {Santhanam}(1990)}]{david}%
  \BibitemOpen
  \bibfield  {author} {\bibinfo {author} {\bibfnamefont {A.}~\bibnamefont
  {Davidson}}\ and\ \bibinfo {author} {\bibfnamefont {P.}~\bibnamefont
  {Santhanam}},\ }\bibfield  {title} {\bibinfo {title} {Quantum rotors,
  phase-slip centers, and the {C}oulomb blockade},\ }\href@noop {} {\bibfield
  {journal} {\bibinfo  {journal} {Phys. Lett. A}\ }\textbf {\bibinfo {volume}
  {149}},\ \bibinfo {pages} {476} (\bibinfo {year} {1990})}\BibitemShut
  {NoStop}%
\bibitem [{\citenamefont {Lifshitz}\ and\ \citenamefont
  {Pitaevskii}(1980)}]{LLstPh2}%
  \BibitemOpen
  \bibfield  {author} {\bibinfo {author} {\bibfnamefont {E.~M.}\ \bibnamefont
  {Lifshitz}}\ and\ \bibinfo {author} {\bibfnamefont {L.~P.}\ \bibnamefont
  {Pitaevskii}},\ }\href@noop {} {\emph {\bibinfo {title} {Statistical physics.
  Part 2}}}\ (\bibinfo  {publisher} {Pergamon Press},\ \bibinfo {year}
  {1980})\BibitemShut {NoStop}%
\bibitem [{\citenamefont {Ziman}(1972)}]{ziman}%
  \BibitemOpen
  \bibfield  {author} {\bibinfo {author} {\bibfnamefont {J.}~\bibnamefont
  {Ziman}},\ }\href@noop {} {\emph {\bibinfo {title} {Principles of the theory
  of solids}}},\ \bibinfo {edition} {2nd}\ ed.\ (\bibinfo  {publisher}
  {Cambridge University Press},\ \bibinfo {year} {1972})\BibitemShut {NoStop}%
\bibitem [{\citenamefont {Widom}\ \emph {et~al.}(1984)\citenamefont {Widom},
  \citenamefont {Megaloudis}, \citenamefont {Clark}, \citenamefont {Mutton},
  \citenamefont {Prance},\ and\ \citenamefont {Prance}}]{widom2}%
  \BibitemOpen
  \bibfield  {author} {\bibinfo {author} {\bibfnamefont {A.}~\bibnamefont
  {Widom}}, \bibinfo {author} {\bibfnamefont {G.}~\bibnamefont {Megaloudis}},
  \bibinfo {author} {\bibfnamefont {T.~D.}\ \bibnamefont {Clark}}, \bibinfo
  {author} {\bibfnamefont {J.~E.}\ \bibnamefont {Mutton}}, \bibinfo {author}
  {\bibfnamefont {R.~J.}\ \bibnamefont {Prance}},\ and\ \bibinfo {author}
  {\bibfnamefont {H.}~\bibnamefont {Prance}},\ }\bibfield  {title} {\bibinfo
  {title} {The {J}osephson pendulum as a nonlinear capacitor},\ }\href@noop {}
  {\bibfield  {journal} {\bibinfo  {journal} {J. Low Temp. Phys.}\ }\textbf
  {\bibinfo {volume} {57}},\ \bibinfo {pages} {651} (\bibinfo {year}
  {1984})}\BibitemShut {NoStop}%
\bibitem [{\citenamefont {Averin}\ \emph {et~al.}(1985)\citenamefont {Averin},
  \citenamefont {Zorin},\ and\ \citenamefont {Likharev}}]{AverLikh}%
  \BibitemOpen
  \bibfield  {author} {\bibinfo {author} {\bibfnamefont {D.}~\bibnamefont
  {Averin}}, \bibinfo {author} {\bibfnamefont {A.}~\bibnamefont {Zorin}},\ and\
  \bibinfo {author} {\bibfnamefont {K.}~\bibnamefont {Likharev}},\ }\bibfield
  {title} {\bibinfo {title} {Bloch oscillations in small {J}osephson
  junctions},\ }\href@noop {} {\bibfield  {journal} {\bibinfo  {journal} {Zh.
  Eksp. Teor. Fiz.}\ }\textbf {\bibinfo {volume} {61}},\ \bibinfo {pages} {407}
  (\bibinfo {year} {1985})},\ \bibinfo {note} {[Sov. Phys.--JETP, {\bf 88},
  692--703 (1985)]}\BibitemShut {NoStop}%
\bibitem [{\citenamefont {Likharev}\ and\ \citenamefont {Zorin}(1985)}]{Likh}%
  \BibitemOpen
  \bibfield  {author} {\bibinfo {author} {\bibfnamefont {K.~K.}\ \bibnamefont
  {Likharev}}\ and\ \bibinfo {author} {\bibfnamefont {A.~B.}\ \bibnamefont
  {Zorin}},\ }\bibfield  {title} {\bibinfo {title} {Theory of the {B}loch-wave
  oscillations in small {J}osephson junctions},\ }\href@noop {} {\bibfield
  {journal} {\bibinfo  {journal} {J. Low Temp. Phys.}\ }\textbf {\bibinfo
  {volume} {59}},\ \bibinfo {pages} {347} (\bibinfo {year} {1985})}\BibitemShut
  {NoStop}%
\bibitem [{\citenamefont {Sch\"on}\ and\ \citenamefont {Zaikin}(1990)}]{Schon}%
  \BibitemOpen
  \bibfield  {author} {\bibinfo {author} {\bibfnamefont {G.}~\bibnamefont
  {Sch\"on}}\ and\ \bibinfo {author} {\bibfnamefont {A.}~\bibnamefont
  {Zaikin}},\ }\bibfield  {title} {\bibinfo {title} {Quantum coherent effects,
  phase transitions, and the dissipative dynamics of ultra small tunnel
  junctions},\ }\href@noop {} {\bibfield  {journal} {\bibinfo  {journal} {Phys.
  Rep.}\ }\textbf {\bibinfo {volume} {198}},\ \bibinfo {pages} {237} (\bibinfo
  {year} {1990})}\BibitemShut {NoStop}%
\bibitem [{\citenamefont {Penttil\"a}\ \emph {et~al.}(1999)\citenamefont
  {Penttil\"a}, \citenamefont {Parts}, \citenamefont {Hakonen}, \citenamefont
  {Paalanen},\ and\ \citenamefont {Sonin}}]{SI}%
  \BibitemOpen
  \bibfield  {author} {\bibinfo {author} {\bibfnamefont {J.~S.}\ \bibnamefont
  {Penttil\"a}}, \bibinfo {author} {\bibfnamefont {U.}~\bibnamefont {Parts}},
  \bibinfo {author} {\bibfnamefont {P.~J.}\ \bibnamefont {Hakonen}}, \bibinfo
  {author} {\bibfnamefont {M.~A.}\ \bibnamefont {Paalanen}},\ and\ \bibinfo
  {author} {\bibfnamefont {E.~B.}\ \bibnamefont {Sonin}},\ }\bibfield  {title}
  {\bibinfo {title} {``{S}uperconductor-insulator transition'' in a single
  {J}osephson junction},\ }\href {https://doi.org/10.1103/PhysRevLett.82.1004}
  {\bibfield  {journal} {\bibinfo  {journal} {Phys. Rev. Lett.}\ }\textbf
  {\bibinfo {volume} {82}},\ \bibinfo {pages} {1004} (\bibinfo {year}
  {1999})}\BibitemShut {NoStop}%
\bibitem [{\citenamefont {Penttil{\"a}}\ \emph {et~al.}(2001)\citenamefont
  {Penttil{\"a}}, \citenamefont {Hakonen}, \citenamefont {Sonin},\ and\
  \citenamefont {Paalanen}}]{Penttila2001}%
  \BibitemOpen
  \bibfield  {author} {\bibinfo {author} {\bibfnamefont {J.~S.}\ \bibnamefont
  {Penttil{\"a}}}, \bibinfo {author} {\bibfnamefont {P.~J.}\ \bibnamefont
  {Hakonen}}, \bibinfo {author} {\bibfnamefont {E.~B.}\ \bibnamefont {Sonin}},\
  and\ \bibinfo {author} {\bibfnamefont {M.~A.}\ \bibnamefont {Paalanen}},\
  }\bibfield  {title} {\bibinfo {title} {Experiments on dissipative dynamics of
  single {J}osephson junctions},\ }\href
  {https://doi.org/10.1023/A:1012971500694} {\bibfield  {journal} {\bibinfo
  {journal} {J. Low Temp. Phys.}\ }\textbf {\bibinfo {volume} {125}},\ \bibinfo
  {pages} {89} (\bibinfo {year} {2001})}\BibitemShut {NoStop}%
\bibitem [{\citenamefont {Caldeira}\ and\ \citenamefont {Leggett}(1983)}]{CL}%
  \BibitemOpen
  \bibfield  {author} {\bibinfo {author} {\bibfnamefont {A.}~\bibnamefont
  {Caldeira}}\ and\ \bibinfo {author} {\bibfnamefont {A.}~\bibnamefont
  {Leggett}},\ }\bibfield  {title} {\bibinfo {title} {Quantum tunneling in a
  dissipative system},\ }\href@noop {} {\bibfield  {journal} {\bibinfo
  {journal} {Ann. Phys. (N.Y.)}\ }\textbf {\bibinfo {volume} {149}},\ \bibinfo
  {pages} {374} (\bibinfo {year} {1983})}\BibitemShut {NoStop}%
\bibitem [{\citenamefont {Tinkham}(1996)}]{Tin}%
  \BibitemOpen
  \bibfield  {author} {\bibinfo {author} {\bibfnamefont {M.}~\bibnamefont
  {Tinkham}},\ }\href@noop {} {\emph {\bibinfo {title} {Introduction to
  superconductivity}}},\ \bibinfo {edition} {2nd}\ ed.\ (\bibinfo  {publisher}
  {McGrow-Hill},\ \bibinfo {year} {1996})\BibitemShut {NoStop}%
\bibitem [{\citenamefont {Mott}(1990)}]{Mott}%
  \BibitemOpen
  \bibfield  {author} {\bibinfo {author} {\bibfnamefont {N.}~\bibnamefont
  {Mott}},\ }\href@noop {} {\emph {\bibinfo {title} {Metal-Insulator
  Transitions}}},\ \bibinfo {edition} {2nd}\ ed.\ (\bibinfo  {publisher} {CRC
  Press},\ \bibinfo {year} {1990})\BibitemShut {NoStop}%
\bibitem [{\citenamefont {Herrero}\ and\ \citenamefont {Zaikin}(2002)}]{Zaik}%
  \BibitemOpen
  \bibfield  {author} {\bibinfo {author} {\bibfnamefont {C.~P.}\ \bibnamefont
  {Herrero}}\ and\ \bibinfo {author} {\bibfnamefont {A.~D.}\ \bibnamefont
  {Zaikin}},\ }\bibfield  {title} {\bibinfo {title} {Superconductor-insulator
  quantum phase transition in a single {J}osephson junction},\ }\href
  {https://doi.org/10.1103/PhysRevB.65.104516} {\bibfield  {journal} {\bibinfo
  {journal} {Phys. Rev. B}\ }\textbf {\bibinfo {volume} {65}},\ \bibinfo
  {pages} {104516} (\bibinfo {year} {2002})}\BibitemShut {NoStop}%
\bibitem [{\citenamefont {Rogovin}\ and\ \citenamefont {Nagel}(1982)}]{Rogov}%
  \BibitemOpen
  \bibfield  {author} {\bibinfo {author} {\bibfnamefont {D.}~\bibnamefont
  {Rogovin}}\ and\ \bibinfo {author} {\bibfnamefont {J.}~\bibnamefont
  {Nagel}},\ }\bibfield  {title} {\bibinfo {title} {Quantum theory of the dc
  josephson effect: Static tunneling characteristics of ultrasmall josephson
  junctions},\ }\href {https://doi.org/10.1103/PhysRevB.26.3698} {\bibfield
  {journal} {\bibinfo  {journal} {Phys. Rev. B}\ }\textbf {\bibinfo {volume}
  {26}},\ \bibinfo {pages} {3698} (\bibinfo {year} {1982})}\BibitemShut
  {NoStop}%
\bibitem [{\citenamefont {Mullen}\ \emph {et~al.}(1993)\citenamefont {Mullen},
  \citenamefont {Loss},\ and\ \citenamefont {Stoof}}]{Loss}%
  \BibitemOpen
  \bibfield  {author} {\bibinfo {author} {\bibfnamefont {K.}~\bibnamefont
  {Mullen}}, \bibinfo {author} {\bibfnamefont {D.}~\bibnamefont {Loss}},\ and\
  \bibinfo {author} {\bibfnamefont {H.~T.~C.}\ \bibnamefont {Stoof}},\
  }\bibfield  {title} {\bibinfo {title} {Resonant phenomena in compact and
  extended systems},\ }\href@noop {} {\bibfield  {journal} {\bibinfo  {journal}
  {Phys. Rev. B}\ }\textbf {\bibinfo {volume} {47}},\ \bibinfo {pages} {2689}
  (\bibinfo {year} {1993})}\BibitemShut {NoStop}%
\bibitem [{\citenamefont {Ben-Jacob}\ and\ \citenamefont
  {Gefen}(1985)}]{GefenJ}%
  \BibitemOpen
  \bibfield  {author} {\bibinfo {author} {\bibfnamefont {E.}~\bibnamefont
  {Ben-Jacob}}\ and\ \bibinfo {author} {\bibfnamefont {Y.}~\bibnamefont
  {Gefen}},\ }\bibfield  {title} {\bibinfo {title} {New quantum oscillations in
  current driven small junctions},\ }\href@noop {} {\bibfield  {journal}
  {\bibinfo  {journal} {Phys. Lett. A}\ }\textbf {\bibinfo {volume} {108}},\
  \bibinfo {pages} {289} (\bibinfo {year} {1985})}\BibitemShut {NoStop}%
\bibitem [{\citenamefont {Zwerger}\ \emph {et~al.}(1986)\citenamefont
  {Zwerger}, \citenamefont {Dorsey},\ and\ \citenamefont {Fisher}}]{Zwerg}%
  \BibitemOpen
  \bibfield  {author} {\bibinfo {author} {\bibfnamefont {W.}~\bibnamefont
  {Zwerger}}, \bibinfo {author} {\bibfnamefont {A.~T.}\ \bibnamefont
  {Dorsey}},\ and\ \bibinfo {author} {\bibfnamefont {M.~P.~A.}\ \bibnamefont
  {Fisher}},\ }\bibfield  {title} {\bibinfo {title} {Effects of the phase
  periodicity on the quantum dynamics of a resistively shunted {J}osephson
  junction},\ }\href@noop {} {\bibfield  {journal} {\bibinfo  {journal} {Phys.
  Rev. B}\ }\textbf {\bibinfo {volume} {34}},\ \bibinfo {pages} {6518}
  (\bibinfo {year} {1986})}\BibitemShut {NoStop}%
\bibitem [{\citenamefont {Apenko}(1989)}]{Apen}%
  \BibitemOpen
  \bibfield  {author} {\bibinfo {author} {\bibfnamefont {S.}~\bibnamefont
  {Apenko}},\ }\bibfield  {title} {\bibinfo {title} {Environment-induced
  decompactification of phase in {J}osephson junctions},\ }\href@noop {}
  {\bibfield  {journal} {\bibinfo  {journal} {Phys. Lett. A}\ }\textbf
  {\bibinfo {volume} {142}},\ \bibinfo {pages} {277} (\bibinfo {year}
  {1989})}\BibitemShut {NoStop}%
\bibitem [{\citenamefont {Morel}\ and\ \citenamefont {Mora}(2021)}]{morel}%
  \BibitemOpen
  \bibfield  {author} {\bibinfo {author} {\bibfnamefont {T.}~\bibnamefont
  {Morel}}\ and\ \bibinfo {author} {\bibfnamefont {C.}~\bibnamefont {Mora}},\
  }\bibfield  {title} {\bibinfo {title} {Double-periodic {J}osephson junctions
  in a quantum dissipative environment},\ }\href
  {https://doi.org/10.1103/PhysRevB.104.245417} {\bibfield  {journal} {\bibinfo
   {journal} {Phys. Rev. B}\ }\textbf {\bibinfo {volume} {104}},\ \bibinfo
  {pages} {245417} (\bibinfo {year} {2021})}\BibitemShut {NoStop}%
\bibitem [{\citenamefont {Zaikin}\ and\ \citenamefont {Golubev}(2019)}]{golub}%
  \BibitemOpen
  \bibfield  {author} {\bibinfo {author} {\bibfnamefont {A.~D.}\ \bibnamefont
  {Zaikin}}\ and\ \bibinfo {author} {\bibfnamefont {D.~S.}\ \bibnamefont
  {Golubev}},\ }\href@noop {} {\emph {\bibinfo {title} {Dissipative Quantum
  Mechanics of Nanostructures. Electron Transport, Fluctuations, and
  Interactions}}}\ (\bibinfo  {publisher} {Taylor \& {F}rancis},\ \bibinfo
  {year} {2019})\BibitemShut {NoStop}%
\bibitem [{\citenamefont {Chen}\ \emph {et~al.}(1988)\citenamefont {Chen},
  \citenamefont {Fisher},\ and\ \citenamefont {leggett}}]{Legg}%
  \BibitemOpen
  \bibfield  {author} {\bibinfo {author} {\bibfnamefont {Y.~C.}\ \bibnamefont
  {Chen}}, \bibinfo {author} {\bibfnamefont {M.~P.~A.}\ \bibnamefont
  {Fisher}},\ and\ \bibinfo {author} {\bibfnamefont {A.~J.}\ \bibnamefont
  {leggett}},\ }\bibfield  {title} {\bibinfo {title} {The return of a
  hysteretic {J}osephson junction to the zero-voltage state: {\it I --V}
  characteristic and quantum retrapping},\ }\href@noop {} {\bibfield  {journal}
  {\bibinfo  {journal} {J. Appl. Phys.}\ }\textbf {\bibinfo {volume} {64}},\
  \bibinfo {pages} {3119} (\bibinfo {year} {1988})}\BibitemShut {NoStop}%
\bibitem [{\citenamefont {Averin}\ and\ \citenamefont {Bruder}(2018)}]{averin}%
  \BibitemOpen
  \bibfield  {author} {\bibinfo {author} {\bibfnamefont {D.~V.}\ \bibnamefont
  {Averin}}\ and\ \bibinfo {author} {\bibfnamefont {C.}~\bibnamefont
  {Bruder}},\ }\bibfield  {title} {\bibinfo {title} {Indistinguishability of
  quantum states and rotation counting},\ }\href@noop {} {\bibfield  {journal}
  {\bibinfo  {journal} {Phys. Rev. B}\ }\textbf {\bibinfo {volume} {98}}
  (\bibinfo {year} {2018})}\BibitemShut {NoStop}%
\end{thebibliography}
%apsrev4-2.bst 2019-01-14 (MD) hand-edited version of apsrev4-1.bst
%Control: key (0)
%Control: author (8) initials jnrlst
%Control: editor formatted (1) identically to author
%Control: production of article title (0) allowed
%Control: page (0) single
%Control: year (1) truncated
%Control: production of eprint (0) enabled

%apsrev4-2.bst 2019-01-14 (MD) hand-edited version of apsrev4-1.bst
%Control: key (0)
%Control: author (8) initials jnrlst
%Control: editor formatted (1) identically to author
%Control: production of article title (0) allowed
%Control: page (0) single
%Control: year (1) truncated
%Control: production of eprint (0) enabled
%

\end{document}